\documentclass[a4paper,12pt]{jaa}
\usepackage[british]{babel}             
\usepackage{newtxtext}                  
\usepackage[slantedGreek]{newtxmath}    
\usepackage[T1]{fontenc}                
\usepackage{graphicx}                   
\usepackage[authoryear,longnamesfirst]{natbib}
\bibpunct{(}{)}{;}{a}{}{,}
\usepackage{hyperref}
\hypersetup{pdfauthor={D. A. Green},
            pdftitle={An updated catalogue of 310 Galactic supernova remnants
                      and their statistical properties},
            pdfkeywords={supernova remnants, catalogues, ISM: general},
            bookmarksnumbered=true}
\usepackage{xcolor}
\hypersetup{colorlinks=true,linkcolor=purple,citecolor=teal,urlcolor=violet}
\usepackage{fix-jaa}
\def\SNR(#1)#2(#3){{G#1$#2$#3}}
\def\SigmaUnit{{\rm W~m^{-2}\,Hz^{-1}\,sr^{-1}}}

\makeatletter
\def\ps@firstpage{
\def\@oddfoot{}
\let\@evenhead\@empty
\def\@oddhead{{\noindent\hskip-6pt\begin{tabular}[t]{l}
\\[-24pt]\fontsize{9}{11}\selectfont\textit{J. Astrophys. Astr.} (2024) \textbf{45}: \the\artcitid\\
\fontsize{9}{11}\selectfont DOI \end{tabular}\hfill \setlength\tabcolsep{0pt} }
}}
\def\ps@myheadings{%
    \def\@evenhead{\small\the\artcitid\ Page \thepage\ of\ \the\lp\hfil\textit{J. Astrophys. Astr.} (2024) \textbf{45}: \the\artcitid}
    \def\@oddhead{\small{\textit{J. Astrophys. Astr.} \textrm{(2024) \textbf{45}: \the\artcitid}\hfil Page\ \thepage\ of\ \the\lp\ \the\artcitid}%
    \let\@mkboth\@gobbletwo
    \let\sectionmark\@gobble
    \let\subsectionmark\@gobble
    \renewcommand\footnoterule{\hskip0pt\rule{1in}{.5pt}}
    }}
\makeatother
\begin{document}\sloppy
\doinum{10.1007/...}
\artcitid{in press}
\volnum{45}
\year{2024}
\setcounter{page}{1}

\title{An updated catalogue of 310 Galactic supernova remnants and their\\
statistical properties}

\author{D.~A.\ Green}
\affilOne{Astrophysics Group, Cavendish Laboratory,
       19 J.~J.~Thomson Avenue, Cambridge CB3 0HE, United Kingdom}

\twocolumn[{

\maketitle

\corres{D.A.Green@mrao.cam.ac.uk}

\msinfo{19 Aug 2024}{?}

\label{firstpage}

%
%
%

\begin{abstract}
A revised catalogue of 310 Galactic supernova remnants (SNRs) is
presented, along with some statistics of their properties. 21 SNRs have
been added to the catalogue since the previous published version from
2019, and 5 entries have been removed, as they have been identified as
{\sc H\,ii} regions. Also discussed are some basics statistics of the
remnants in the catalogue, the selection effects that apply to the
identification of Galactic SNRs and their consequences.
\end{abstract}


}]

\section{Introduction}\label{s:intro}

This paper presents the latest version of  a catalogue of Galactic
supernova remnants (SNRs) which I have compiled for several decades.
Previous versions have been published in \citep{1984MNRAS.209..449G,
1988Ap&SS.148....3G, 1991PASP..103..209G, 2002ISAA....5.....S,
2004BASI...32..335G, 2009BASI...37...45G, 2014BASI...42...47G,
2019JApA...40...36G}. In addition, more detailed web-based versions of
the catalogue have been produced since 1995 which either correspond to
one of the published catalogues, or are an intermediate revision. This
version of the catalogue contains 310 entries. Section~\ref{s:cat} gives
the details of the entries in the catalogue, and Section~\ref{s:new}
discusses the entries added or removed from the catalogue since the last
published version \citep{2019JApA...40...36G}. Section~\ref{s:discuss}
discusses some statistics of the remnants in the current catalogue, the
selection effects that apply to the identification of Galactic SNRs, and
their consequences.

\section{The catalogue format}\label{s:cat}

This catalogue is based on the literature published up to the end of
2023, and contains 310 entries. For each SNR in the catalogue the
following parameters are given.
\begin{itemize}
\item {\bf Galactic Coordinates} of the remnant. These are quoted to a
tenth of a degree, as is conventional. In this catalogue additional
leading zeros are not used. These are generally taken from the Galactic
coordinate based name used for the remnant in the literature. It should
be noted that when these names were first defined, they may not follow
the IAU recommendation\altfoonoterule\footnote{See:
{\scriptsize\url{http://cdsweb.u-strasbg.fr/Dic/iau-spec.htx}}.} that
coordinates should be truncated, not rounded to construct such names.
\item {\bf Right Ascension} and {\bf Declination} of J2000.0 equatorial
coordinates of the source centroid, for which the accuracy of the quoted
values depends on the size of the remnant. For small remnants they are
to the nearest few seconds of time and the nearest minute of arc
respectively, whereas for larger remnants they are rounded to coarser
values, but are in every case sufficient to specify a point within the
boundary of the remnant. These coordinates are usually deduced from
radio images rather than from X-ray or optical observations.
\item {\bf Angular Size} of the remnant, in arcminutes. This is usually
taken from the highest resolution radio image available. The boundary of
most remnants approximates reasonably well to either a circle or to an
ellipse. A single value is quoted for the angular size of the more
nearly circular remnants, which is the diameter of a circle with an area
equal to that of the remnant. For more elongated remnants the product of
two values is given, which are the major and minor diameters of the
remnant boundary modelled as an ellipse. In a small number of cases an
ellipse is not a good description of the boundary of the object (which
will be noted in the description of the object given in its catalogue
entry), although an angular size is still quoted for information. For
`filled-centre' type remnants (see below), the size quoted is for the
largest extent of the observed emission, not, as at times has been used
by others, the half-width of the centrally brightened peak.
\item {\bf Type} of the SNR: `S' or `F' if the remnant shows a `shell'
or `filled-centre' structure, or `C' if it shows `composite' (or
`combination') radio structure, with a combination of shell and
filled-centre characteristics. If there is some uncertainty, the type is
given as `S?', `F?' or `C?', or as `?' in several cases where an object
is conventionally regarded as an SNR even though its nature is poorly
known or it is not well-understood. (Note: the term `composite' has been
used, by some authors, in a different sense, to describe remnants with
radio shell and centrally-brightened X-ray emission. An alternative term
used to describe such remnants is `mixed morphology', e.g.\ see
\citealt{1998ApJ...503L.167R}.)
\item {\bf Flux Density} of the remnant at a frequency of 1~GHz, in
jansky, if available. Not all entries have a value, as some new remnants
have not been identified at radio wavelengths, or else the available
radio observations are not good enough to provide a reliable flux
density. These values are  {\sl not} measured values, but are instead
derived from the observed radio spectrum of the source. The frequency of
1~GHz is chosen because flux density measurements are usually available
at both higher and lower frequencies. Some young remnants -- notably
\SNR(111.7)-(2.1) ($=$Cassiopeia A) and \SNR(184.6)-(5.8) ($=$Crab
Nebula) -- show secular variations in their radio flux density. In the
catalogue the 1-GHz flux densities for \SNR(111.7)-(2.1) and
\SNR(184.6)-(5.8) have been taken from \citep{2017ApJS..230....7P}, for
an epoch of 2016. Results from the primary literature should be used for
any detailed quantitative studies of the radio flux densities of these
and other remnants.
\item {\bf Spectral Index} of the integrated radio emission from the
remnant, $\alpha$ (here defined in the sense, $S \propto \nu^{-\alpha}$,
where $S$ is the flux density at frequency $\nu$). This is either a
value that is quoted in the literature, or one deduced from the
available integrated flux densities of the remnant. For several SNRs a
simple power law is not adequate to describe their radio spectra, either
because there is evidence that the integrated spectrum is curved or the
spectral index varies across the face of the remnant. In these cases the
spectral index is given as `varies' (refer to the description of the
remnant and appropriate references in the detailed catalogue entry for
more information). In some cases, for example where the remnant is
highly confused with thermal emission, the spectral index is given as
`?' since no value can be deduced with any confidence. These spectral
indices have a very wide range of quality, and the primary literature
should be consulted for any detailed study of the radio spectral indices
of Galactic remnants.
\item {\bf Other Names} that are commonly used for the remnant. Note
that these are given in parentheses if the remnant is only a part of the
source. For some well known remnants -- e.g.\ \SNR(184.6)-(5.8), the
Crab Nebula -- not all common names are given.
\end{itemize}
A summary of the data available for all 310 remnants in the catalogue is
given in Table~\ref{t:snrcat}.

A more detailed version of the catalogue is available on the
web\footnote{See:
{\scriptsize\url{https://www.mrao.cam.ac.uk/surveys/snrs/}}.}. In
addition to the basic parameters which are given in
Table~\ref{t:snrcat}, the detailed version of the catalogue contains the
following additional information.
(i) Notes on the remnant. For example, if other Galactic coordinates
have at times been used to label it (usually before good observations
have revealed the full extent of the object, but sometimes in error); if
the SNR is thought to be the remnant of a historical SN.
(ii) Short descriptions of the observed structure/properties of the
remnant at radio, optical and X-ray wavelengths, as appropriate from
available observations.
(iii) Comments on distance determinations, and any point sources or
pulsars in or near the object (although they may not necessarily be
related to the remnant).
(iv) References to observations are given for each remnant, complete
with journal, volume, page, and a short description of what information
each paper contains (e.g.\ for radio observations these generally
include the telescopes used, the observing frequencies and resolutions,
together with any flux density determinations). These references are
{\sl not} complete, but cover recent and representative observations of
the remnant that are available, and should themselves include references
to earlier work. These references are from the published literature up
to the end of 2023, from `supernova remnant' abstract/title/keyword
searches on the `Astrophysics Data System'\footnote{See:
{\scriptsize\url{https://ui.adsabs.harvard.edu/}}.} (ADS).

The detailed version of the catalogue is available in pdf format for
downloading and printing, or as web pages, including a page for each
individual remnant. The web pages for each remnant include links to
ADS for each of
the over three thousand references that are included in the detailed
listings, and links are also provided for either title/abstract/keyword
or full-text searches at ADS by the SNR name(s).

Some of the parameters included in the catalogue are themselves of
variable quality. For example, the radio flux density of each remnant at
1~GHz is generally obtained from several radio observations over a range
of frequencies, both above and below 1~GHz, so is of good quality.
However, there are 27 remnants -- often those which have been identified
at other than radio wavelengths -- for which no reliable radio flux
density is yet available, because they have either not been detected or
well observed in the radio. Although the detailed version of the
catalogue contains notes on distances for many remnants reported in the
literature, these are highly variable in terms of reliability and
accuracy. Consequently the distances given within the detailed version
of catalogue should be used with caution in any statistical studies, and
reference should be made to the primary literature cited in the detailed
catalogue. For SNR distances, see also \citet{2022ApJ...940...63R}, who
provide a compilation of distances for 215 Galactic SNRs (although not
all of these are included in the catalogue), which are discussed briefly
in Section~\ref{s:SigmaD}.

The detailed version of the catalogue also contains notes both on those
objects no longer thought to be SNRs, and on the many possible and
probable remnants that have been reported in the literature (including
possible large, old remnants, seen from radio continuum, X-ray or {\sc
H\,i} observations). In this revision of the catalogue the detailed
version now also includes a list of many of these proposed remnants. See
also Section \ref{s:pp} below, for discussion of some of the recently
proposed remnants.

It should be noted that the catalogue is far from homogeneous. Although
many remnants, or possible remnants, were first identified from
wide-area radio surveys, there are many others that have been observed
with diverse observational parameters. This makes uniform criteria for
inclusion in the main catalogue difficult, and the dividing
line between entries in the main catalogue and in the list of
possible and probable remnants (see Section~\ref{s:pp}) is subjective.

For an alternative catalogue of high-energy observations of Galactic
SNRs see \citet{2012AdSpR..49.1313F}\footnote{See:
{\scriptsize\url{http://snrcat.physics.umanitoba.ca/}} for
the current version.}.

\section{Additions and Removals}\label{s:new}

Since the last published version of the catalogue
\citep{2019JApA...40...36G}, the following SNRs have been
added.

\begin{itemize}
\item \citet{2019PASA...36...48H} identify several new SNRs, from
improved radio observations, which had previously been suggested as
candidate remnants by \citet{1990ApJ...364..187G},
\citet{1994MNRAS.270..847G}, \citet{1995MNRAS.277...36D},
\citet{1997MNRAS.287..722D}, \citet{1996A&AS..118..329W},
\citet{2006ApJ...639L..25B} and \citet{2008ApJ...681..320R}, namely:
\SNR(3.1)-(0.6), \SNR(7.5)-(1.7), \SNR(13.1)-(0.5), \SNR(15.5)-(0.1),
\SNR(28.3)+(0.2), \SNR(28.7)-(0.4), \SNR(345.1)-(0.2),
\SNR(345.1)+(0.2), \SNR(348.8)+(1.1), \SNR(353.3)-(1.1) and
\SNR(359.2)-(1.1). (Another of the SNRs identified by
\citeauthor{2019PASA...36...48H} has been included in the catalogue, as
\SNR(9.7)-(0.0), since 2006.)
\item \SNR(21.8)-(3.0) identified from radio and other observations by
\citet{2020MNRAS.493.2188G}.
\item \SNR(107.0)+(9.0), a large ring of optical filaments noted by
\citet{2020MNRAS.498.5194F}, which was subsequently studied at radio
wavelengths by \citet{2021A&A...655A..10R}.
\item \SNR(249.5)+(24.5), a large (\hbox{$\approx 4^\circ$}) shell remnant
found by \citet{2021A&A...648A..30B} from eROSITA X-ray and other
observations.
\item \SNR(17.8)+(16.7), a remnant identified from its non-thermal
radio emission by \citet{2022MNRAS.510.2920A}.
\item \citet{2023A&A...671A.145D} confirmed three candidate remnants as
SNRs, from radio observations, including polarisation (all three of
which had been suggested as possible SNRs by
\citealt{2006AJ....131.2525H}). Two of these -- \SNR(28.3)+(0.2) and
\SNR(28.7)-(0.4) -- had already been added to the 2022 December web
version of catalogue following the observations by
\citet{2019PASA...36...48H}. The third, \SNR(29.3)+(0.1), has been added
to this version of the catalogue.
\item A large (\hbox{$\approx ~4^\circ$}), high latitude SNR,
\SNR(116.6)-(26.1), which was first identified as a possible remnant by
\citet{2021MNRAS.507..971C} from X-ray observations, and subsequently
confirmed from radio and optical observations
\citep{2022MNRAS.513L..83C, 2022MNRAS.515..339P}.
\item \SNR(189.6)+(3.3), a faint SNR overlapping \SNR(189.1)+(3.0)
(=IC443) -- first suggested by \citet{1994A&A...284..573A} from ROSAT
X-ray observations -- following improved X-ray eROSITA observations by
\citet{2023A&A...680A..83C}.
\item \citet{2022SCPMA..6529705G} presented radio observations of two
new large SNRs, \SNR(203.1)+(6.6) and \SNR(206.7)+(5.9). These had
previously been reported as possible SNRs by both
\citet{2002nsps.conf....1R} and \citet{2005ASPC..343..286S}.
\item One of the candidate remnants reported by \citet{1997MNRAS.287..722D},
\SNR(288.8)-(6.3), from radio observations has been confirmed as a
large, faint SNR from improved radio observations by
\citet{2023AJ....166..149F}.
\end{itemize}
Note that these include several new remnants identified at
high Galactic latitudes. See further discussion in Section~\ref{s:se}.

In this version of the catalogue 5 objects previously listed as SNRs
have been removed.
\begin{itemize}
\item \SNR(11.1)-(1.0) and \SNR(16.4)-(0.5), which
\citet{2019A&A...623A.105G} identified as {\sc H\,ii} regions rather
than SNRs. (\citeauthor{2019A&A...623A.105G} also identified
\SNR(20.4)+(0.1) as an {\sc H\,ii} region, which had been removed from
the catalogue in 2004),
\item \SNR(8.3)-(0.0), \SNR(10.5)-(0.0) and \SNR(14.3)+(0.1), which
\citet{2021A&A...651A..86D} identified as {\sc H\,ii} regions rather
than SNRs. (\citeauthor{2021A&A...651A..86D} also identified
\SNR(11.1)-(1.0) as an {\sc H\,ii} region.)
\end{itemize}

\section{Discussion}\label{s:discuss}

\subsection{Possible and Probable SNRs}\label{s:pp}

As noted in Section~\ref{s:cat}, the detailed version of the catalogue
includes notes on many objects that have been reported in the literature
as possible or probable SNRs. Here I discuss briefly
some of the recently proposed SNR candidates.
\begin{itemize}
\item \cite{2021A&A...651A..86D} present radio observations in the
region $358^\circ \le l \le 60^\circ$, $|b| \le 1^\circ$, and identify
157 candidate remnants (about half of which are previously proposed
candidate SNRs). See further discussion of these in Section~\ref{s:se}.
\item \citet{2020PASJ...72L..11S} identifies a small diameter hole in CO
emission as a possible `dark' SNR, a new class of remnant (see also
\citealt{2021Galax...9...13S}).
\item \citet{2021ApJ...918L..33R} propose the faint, fast expanding
optical nebula Pa 30 as the remnant of the historical supernova of
AD1181, rather than \SNR(130.7)+(3.1) (=3C58). See also
\citet{2023ApJ...945L...4F, 2023ApJ...944..120L}. However, unlike known
young SNRs this has not been detected in the radio (for example see the
Canadian Galactic Plane Survey, \citealt{2003AJ....125.3145T}).
\end{itemize}

\begin{figure}
\centerline{\includegraphics[width=8.5cm]{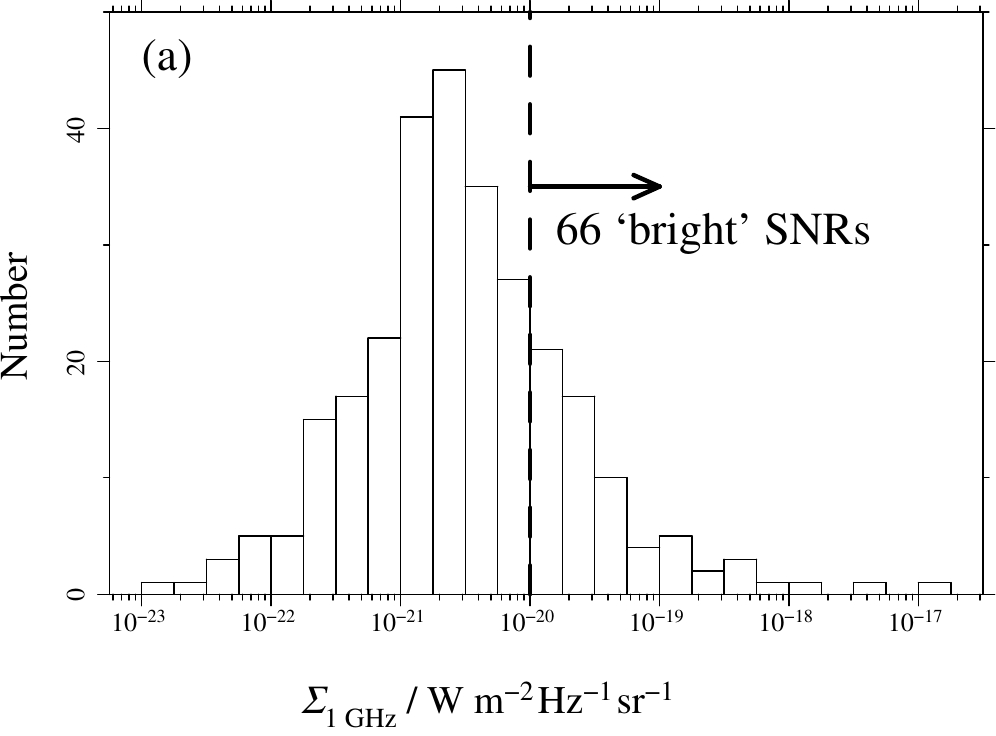}}
\bigskip
\centerline{\includegraphics[width=8.5cm]{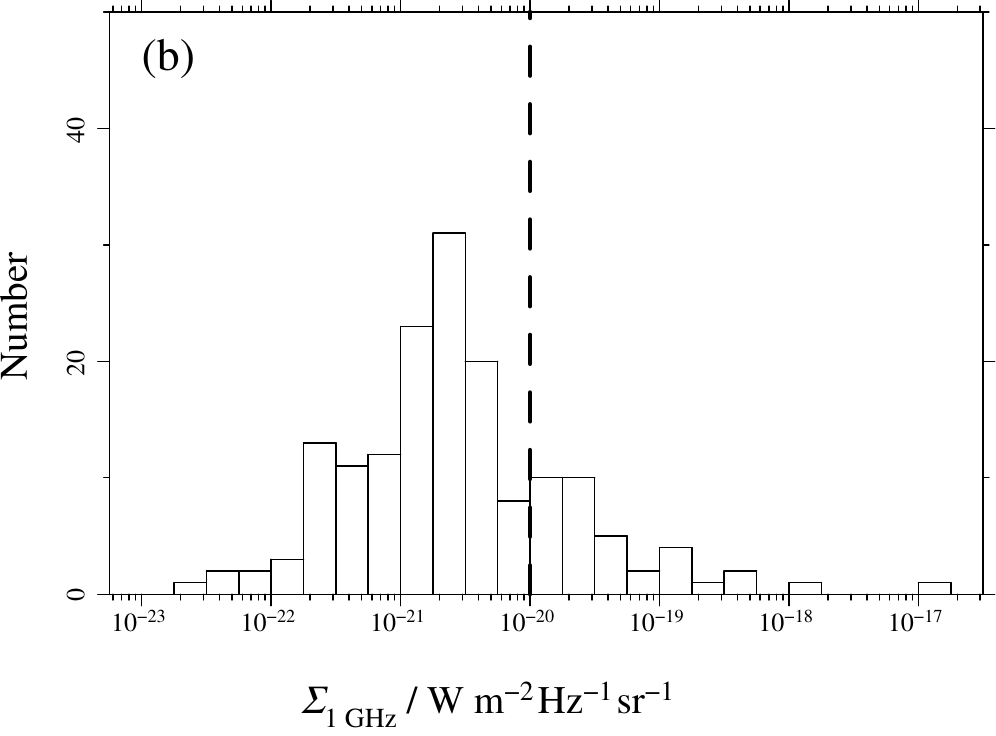}}
\bigskip
\centerline{\includegraphics[width=8.5cm]{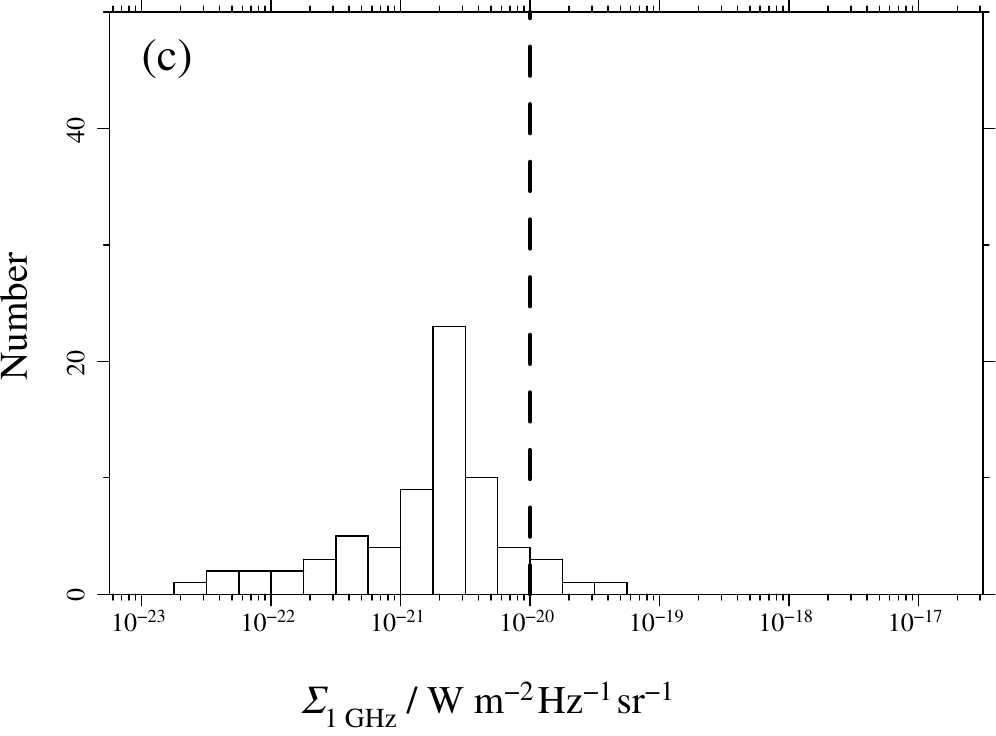}}
\caption{Histograms for the number of SNRs in the current catalogue with
surface brightness at 1~GHz: (a) all remnants; (b) remnants in
the Effelsberg 2.7-GHz Galactic plane survey region (i.e.\ $358^\circ <
l < 240^\circ$, $|b| < 5^\circ$); (c) remnants in the Effelsberg
2.7-GHz Galactic plane survey region added to the catalogue since
1991.\label{f:h-Sigma}}
\end{figure}

\subsection{Some Simple Statistics}\label{s:ss}


In the catalogue there are 23 SNRs which do not have a flux density at
1~GHz and 4 with lower limits. This is because either the remnant has
not been detected at radio wavelengths, or it is poorly defined by
current radio observations, so that their flux density at 1~GHz cannot
be determined with any confidence. So 92\% of the remnants do have a
flux density at 1~GHz in the catalogue. Of the catalogued remnants,
$\approx 46$\% are detected in X-rays, and $\approx 32$\% in the
optical. The smaller proportion of SNR identified in the optical and
X-ray wavebands is due to Galactic absorption, which hampers the
detection of distant remnants.

In this version of the catalogue, 80\% of remnants are classified as
shell (or possible shell) remnants, 12\% are composite (or possible
composite) remnants, and just 3\% are filled-centre (or possible filled
centre) remnants. The types of the remaining remnants are not clear from
current observations (or else they are objects which are conventionally
regarded as SNRs although they do not fit well into any of the
conventional types, e.g.\ CTB80 ($=$\SNR(69.0)+(2.7)), MSH 17$-$3{\em 9}
($=$\SNR(357.7)-(0.1))).

\begin{figure}
\centerline{\includegraphics[width=8.5cm]{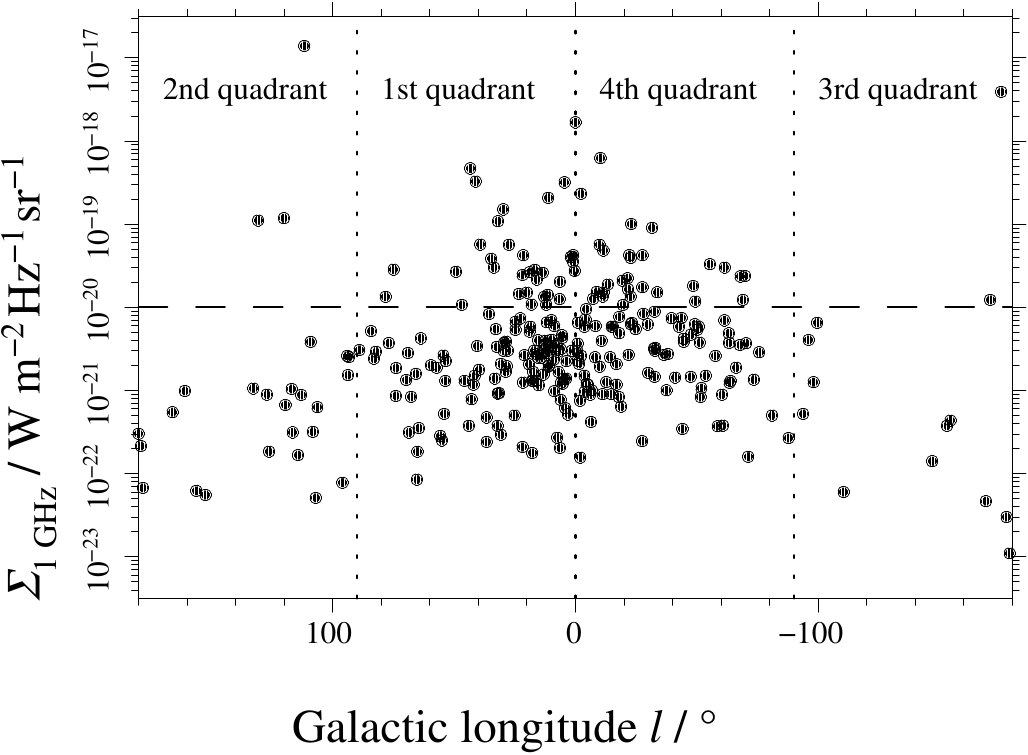}}
\bigskip
\centerline{\includegraphics[width=8.5cm]{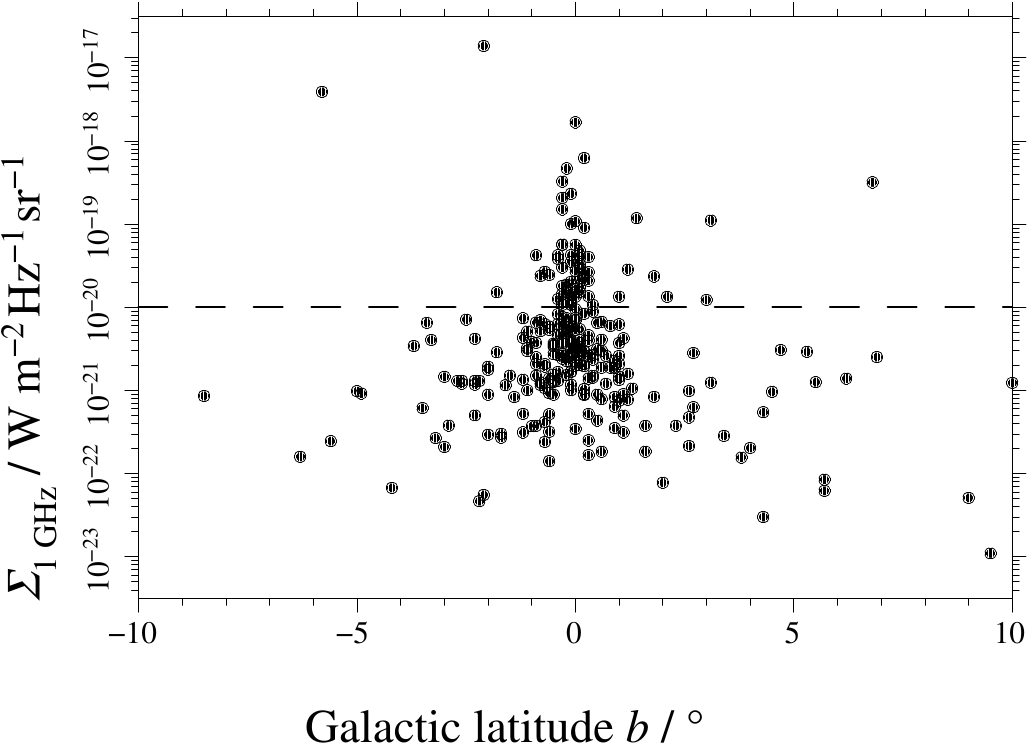}}
\caption{Distribution of SNRs in (top) the $\log{\Sigma}{-}l$ and
(bottom) $\log{\Sigma}{-}b$ planes. All 283 remnants with surface
brightnesses are included in the top plot. There are 5 remnants with
$|b| > 10^\circ$ not included in the bottom plot.\label{f:s-Sigmalb}}
\end{figure}

\begin{figure*}
\centerline{\includegraphics[width=17cm]{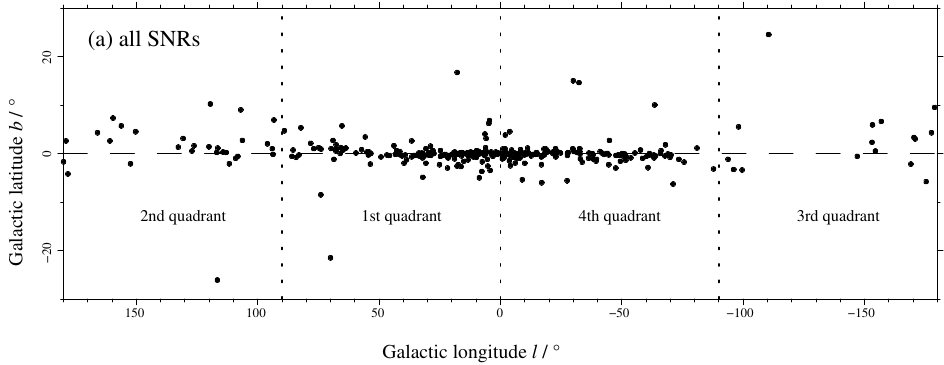}}
\bigskip
\centerline{\includegraphics[width=17cm]{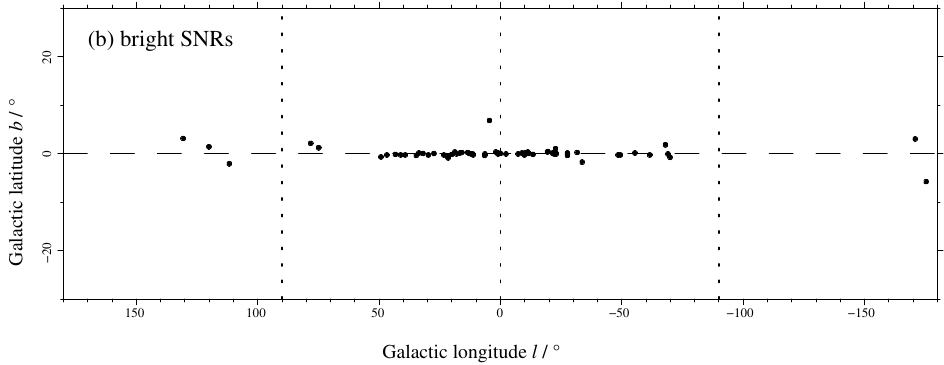}}
\caption{Distribution of SNRs with Galactic latitude and longitude:
(top) all remnants; (bottom) the 66 `bright' remnants. Note
that
the latitude and longitude scales are different.\label{f:l-b}}
\end{figure*}

\begin{figure}
\centerline{\includegraphics[width=8.5cm]{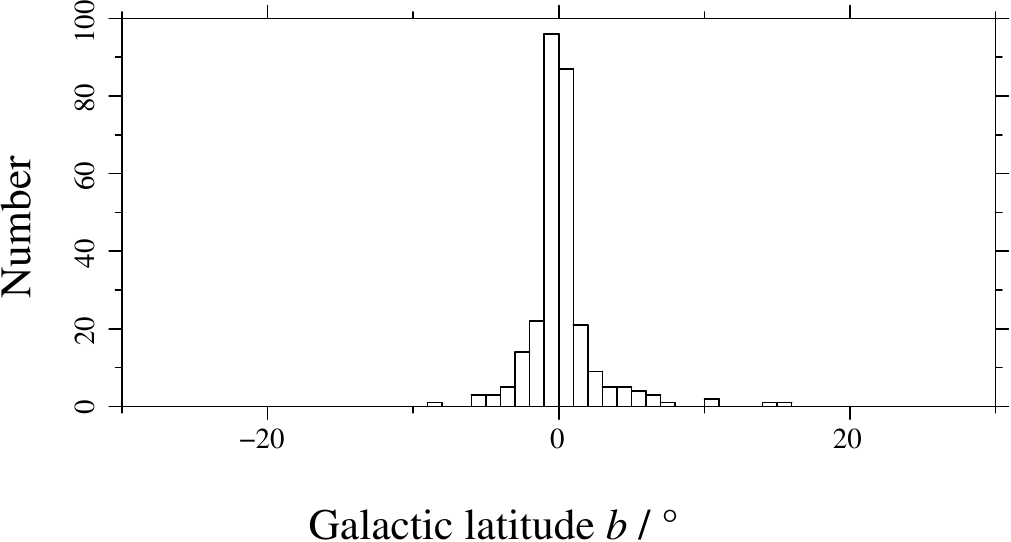}}
\bigskip
\centerline{\includegraphics[width=8.5cm]{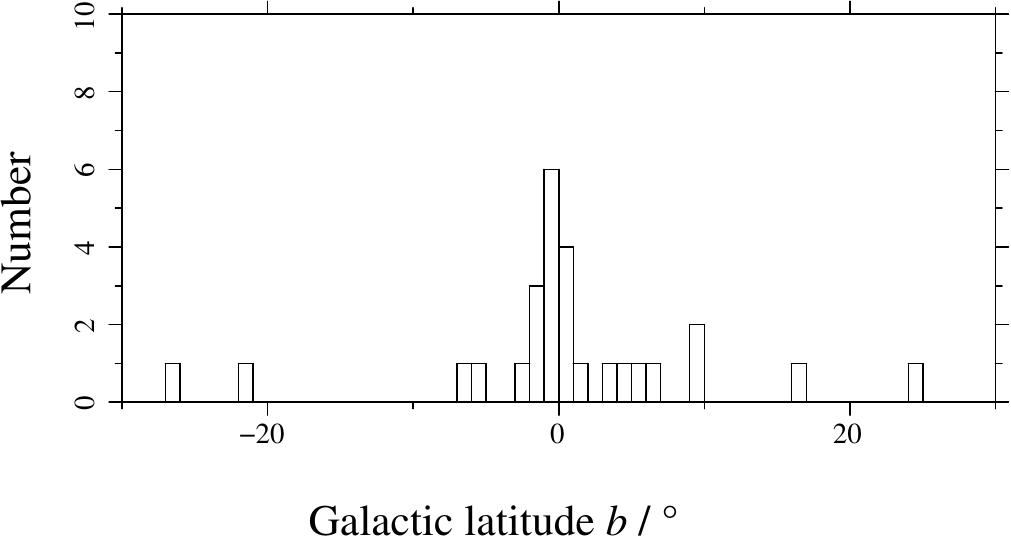}}
\caption{Histograms for the number of SNRs with Galactic latitude:
(top) remnants included in the catalogue before 2017 and (bottom)
remnants included in the catalogue from 2017 or later. Note
that these histograms have different number scales.\label{f:h-b}}
\end{figure}

\subsection{Selection Effects}\label{s:se}

Previously (e.g.\ \citealt{1991PASP..103..209G, 2005MmSAI..76..534G}) I
have discussed the selection effects that apply to the identification of
Galactic SNRs. Although some SNR are identified first at other than
radio wavelengths, most SNRs have been identified first in the radio --
and also many remnants are not detected in the optical or in X-rays due
to obscuration -- and therefore the selection effects for the SNR
catalogue are dominated by those that apply at radio wavelengths. These
are: (i) the difficulty in identifying low surface brightness remnants
the presence of the extended Galactic background, and (ii) the
difficulty in recognising small angular size remnants. In
\citet{2005MmSAI..76..534G} I derived a surface brightness completeness
limit, at 1~GHz, of $\Sigma_{\rm 1~GHz} \approx 10^{-20}$
${\SigmaUnit}$. This is based on the Effelsberg 2.7~GHz survey of a
large part of the Galactic plane (i.e.\ $358^\circ < l < 240^\circ$,
$|b| < 5^\circ$, see: \citealt{1988srim.conf..293R, 1990A&AS...85..633R,
1990A&AS...85..691F}). However, the surface brightness completeness
limit is likely to be somewhat higher close to the Galactic Centre,
where the background Galactic radio emission is brightest, and it is
most difficult to identify faint SNRs.
Note that surface brightess is independent of distance.
This nominal completeness limit
is supported by various searches for SNRs, as no remnants with a surface
brightness above this limit have been added to the published versions of
the catalogue since \citet{2009BASI...37...45G}. This is illustrated in
Fig.~\ref{f:h-Sigma}, which shows histograms of SNRs in the catalogue
with surface brightness: (a) all 283 with a surface brightness; (b)
those in Effelsberg 2.7~GHz survey region, and (c) those in the
Effelsberg 2.7~GHz survey region identified since 1991. In the region
covered by the Effelsberg 2.7~GHz survey there are 172 remnants (i.e.\
Fig.~\ref{f:h-Sigma}b), of which 80 have been added to the catalogue
since 1991 (i.e.\ Fig.~\ref{f:h-Sigma}c). The majority of the remnants
added after 1991 are below the nominal surface brightness limit of
$\approx 10^{-20}$ ${\SigmaUnit}$. The five above the limit are
\SNR(0.3)+(0.0), \SNR(1.0)-(0.1), \SNR(6.5)-(0.4), \SNR(12.8)-(0.0) and
\SNR(18.1)-(0.1), which are all close to the Galactic Centre. Of these
\SNR(1.0)-(0.1) is the brightest, with $\Sigma_{\rm 1~GHz} \approx 3.5
\times 10^{-20}$ $\SigmaUnit$.

This approximate surface brightness limit is consistent with the
properties of the 157 SNR candidates reported by
\citet{2021A&A...651A..86D} that were noted in Section~\ref{s:pp}. Flux
densities at 5.8~GHz are given for many of these candidates, along with
their angular sizes.  None of these candidate has a surface brightness
above $10^{-20}$ ${\SigmaUnit}$, at 1~GHz (assuming a spectral index of
$\alpha = 0.5$).

There are 66 remnants in the current catalogue brighter than the nominal
surface brightness completeness limit of $10^{-20}$ ${\SigmaUnit}$.
Since it is easier to identify SNRs in regions where the background
Galactic emission radio is fainter, i.e.\ away from $l=0^\circ$ or
$b=0^\circ$, the distribution of all SNRs in the catalogue is biased
towards these regions. This is illustrated in Fig.~\ref{f:s-Sigmalb} and
\ref{f:l-b}. Figure~\ref{f:s-Sigmalb} shows the distributions of surface
brightness with Galactic longitude and latitude. These show both that
faintest identified remnants are in the Galactic anti-centre (i.e.\ the
2nd and 3rd Galactic quadrants), and at higher Galactic latitudes, and
also that a larger proportion of SNRs in these regions are faint.
Figure~\ref{f:l-b} shows the Galactic distribution of all catalogued
SNRs and 66 `bright' ones. This shows that distribution of `bright' SNRs
are much more concentrated towards the inner Galactic plane (i.e.\ near
$b=0^\circ$ for the 1st and 4th Galactic quadrants) than that of all
catalogue remnants, which gives a biased view of their distribution due
to the surface brightness selection effect.

It was noted in Section~\ref{s:new} that in recent years several newly
identified SNRs are at relatively high latitude. These include
\SNR(249.5)+(24.5) and \SNR(116.6)-(26.1) first identified from eROSITA X-ray
observations which are not restricted to the Galactic plane. It is away
from the Galactic place where fainter remnants can be more easily
identified, which applies not just at radio wavelengths.
Figure~\ref{f:h-b} shows histograms of SNRs in the catalogue with
Galactic latitude, $b$, for those included in the catalogue in 2017 or
earlier, and those added after 2017. For those catalogued in 2017 or
earlier 17 of 283 (i.e.\ 6\%) have $|b| \ge 5^\circ$, whereas a much
larger proportion of those added after 2017, 10 of 27 (i.e.\ 37\%) have
$|b| > 5^\circ$.

The other selection effect that applies is that small angular size
remnants -- which will be the young but distant SNRs in the Galaxy --
need to be resolved, for their structure to be recognised. Most wide
field radio surveys have not had good enough resolutions to easily
identify small angular size remnants, see \citet{2004BASI...32..335G,
2005MmSAI..76..534G} for previous discussion of this.

It is important to note that the surface brightness selection effect is
more of a problem towards $l=0^\circ$ or $b=0^\circ$. This complicates
studies of the properties of Galactic SNRs that also correlate with
Galactic coordinates. One example is the distribution of Galactic SNRs
with Galactocentric radius, where fainter remnants are easier to
identify in the Galactic anti-centre, at large Galactocentric radius.
\citet{2022ApJ...940...63R} present studies of the distribution of
Galactic SNRs with Galactocentric radius, using a correction factor to
deal with selection effects. However, they assumed this correction
depends on the distance from the Sun, which is not correct for the
surface brightness selection effect, which is independent of distance.
It is more difficult to identify SNRs towards the Galactic centre than
it is for SNRs at the same distance towards the Galactic anti-centre,
where the Galactic background emission is much fainter. Hence
\citet{2022ApJ...940...63R}'s results do not properly take into account
the reality of the observational selection effects. See
Section~\ref{s:gd} for further discussion of studies of the distribution
of SNRs with Galactocentric radius, using a sample of `bright' SNRs to
deal with the surface brightness selection effect. As previously noted,
the surface brightness selection effect also correlates with Galactic
latitude, which  means that faint SNRs are easier to identify a large
$|b|$. Hence SNRs at larger distances above or below the Galactic plane
are easier to identify, as they are in regions with lower Galactic
background emission. \citet{2023ApJS..265...53R} present the
distribution of SNRs with height above/below the Galactic plane, but
without considering the bias due to the surface brightness selection
effect.

\begin{figure}
\centerline{\includegraphics[angle=270,width=8.5cm]{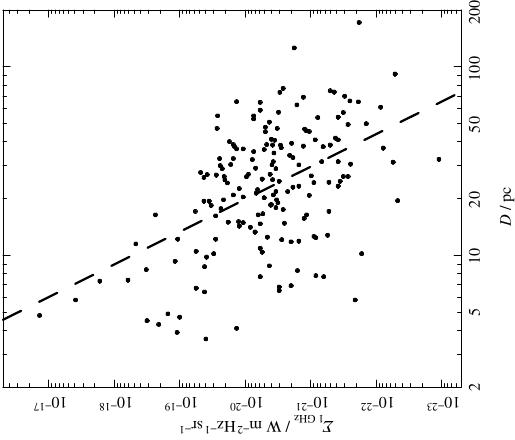}}
\caption{Distribution of 174 remnants with distances, from
\citet{2022ApJ...940...63R}, in the $\log(\Sigma){-}\log(D)$ plane. The
dashed line is a least squares fit minimising deviations in
$\log(D)$.\label{f:s-Sigma+D}}
\end{figure}

\begin{figure}
\centerline{\includegraphics[angle=270,width=8.5cm]{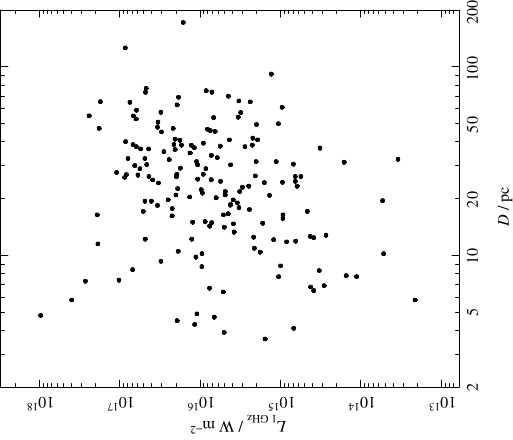}}
\caption{Distribution of 174 remnants with distances, from
\citet{2022ApJ...940...63R}, in the $\log(L){-}\log(D)$
plane.\label{f:s-L+D}}
\end{figure}

\subsection{\texorpdfstring{`$\Sigma{-}D$'}{`Sigma-D'} relation (and
\texorpdfstring{`$L{-}D$'}{`L-D'})}\label{s:SigmaD}

As noted above, \citet{2022ApJ...940...63R} provide a compilation of
distances for 215 Galactic SNRs. 16 of these are not included in the
catalogue, but are included in the possible and probable remnants listed
in the detailed version of the catalogue. Of those in the catalogue,
several do not have radio flux densities at 1~GHz, giving a sample of
174 SNRs with surface brightness ($\Sigma$), distance ($d$) and physical
diameter ($D$, from $d$ and $\theta$). Although the reliability of the
individual distance measurements or estimates are variable, this is a
useful large sample of Galactic SNRs with distances.
Figure~\ref{f:s-Sigma+D} shows the distribution of these 174 remnants in
the $\log(\Sigma){-}\log(D)$ plane, i.e.\ the `$\Sigma{-}D$' diagram,
along with a least square fit straight line fit (which minimises square
deviations in $\log(D)$, which is appropriate if it were to be used to
estimate $D$ from an observed $\Sigma$). This shows that these Galactic
SNRs have a wide spread of properties, and hence (as discussed
previously, e.g.\ \citealt{ 1991PASP..103..209G, 2004BASI...32..335G,
2023ApJS..265...53R}), using the $\Sigma{-}D$ fit to obtain a physical
diameter, $D$, for an individual remnant -- and hence its distance $d$
via its angular size $\theta$ -- is very unreliable.
Figure~\ref{f:s-Sigma+D} shows, for a large range of $\Sigma$, a spread
of about an order of magnitude in $D$ for a given surface brightness. It
should also be remembered that observational selection effects mean that
identification of low surface brightness remnants, and physically small
remnants is difficult, so the range of properties of SNRs to the bottom
and left of the diagram may not be complete. Figure~\ref{f:s-Sigma+D}
does show some correlation between surface brightness and physical
diameter. However, as I have noted previously (e.g.\
\citealt{2005MmSAI..76..534G}), much of the correlation is because
surface brightness is being plotted against physical diameter. Since
$\Sigma \propto S/\theta^2$, and luminosity $L \propto Sd^2$ then
$\Sigma \propto L /(\theta d)^2$, or $\Sigma \propto L/D^2$. Thus the
$\Sigma{-}D$ diagram has a $D^{-2}$ bias compared with a $L-D$ diagram.
Figure~\ref{f:s-L+D} shows the distribution of the sample of 175 SNRs
with distances and flux densities in the $\log(L){-}\log(D)$ plane,
which does not show any obvious correlation.

\subsection{Galactocentric Distribution}\label{s:gd}

As noted in Section~\ref{s:se}, studies of the distribution of SNRs with
Galactocentric radius, $R$, are complicated by the surface brightness
selection effect. An alternative approach is to use a sample of `bright'
SNRs, to mitigate this, and then compare their distribution with
Galactic longitude with that expected from a model. In
\citet{2015MNRAS.454.1517G} I used a sample of 69 SNRs with a surface
brightness, at 1~GHz, above $10^{-20}$ $\SigmaUnit$, and compared these
with a power-low/exponential model distribution of SNRs of the form
$(R/R_\odot)^\alpha \exp( -\beta (R - R_\odot)/R_\odot)$ (as this is the
form used by previous studies, e.g.\ \citealt{1998ApJ...504..761C}). The
best fit parameters produced a distribution that peaked at a smaller
Galactocentric radius than the result from \citet{1998ApJ...504..761C},
and hence had a larger fraction of remnants within the Solar circle
(77\% rather than 49\%).

The current catalogue has 66 SNRs above the surface brightness limit of
$10^{-20}$ $\SigmaUnit$, which is three less than in the sample of
`bright' SNRs used by \citet{2015MNRAS.454.1517G}. This is because: (i)
two objects -- \SNR(20.4)+(0.1) and \SNR(23.6)+(0.3) -- have since been
identified as {\sc H\,ii} regions not SNRs (see
\citealt{2017A&A...605A..58A}), and so have removed from the catalogue,
and (ii) more recent radio observations by \citet{2023MNRAS.524.1396B}
for \SNR(327.4)+(0.4) give a slightly lower flux density for it at
1~GHz, so its surface brightness now falls just below $10^{-20}$
$\SigmaUnit$. However, changing the sample of `bright' SNRs from 69 to
66 does not change the main results of fitting a power-law/exponential
model to the observed distribution of Galactic longitude (i.e.\ the
distribution peaks at a smaller Galactocentric radius with a larger
fraction with the Solar circle than obtained by
\citeauthor{1998ApJ...504..761C}).

\citet{2021MNRAS.504.1536V} present a similar study of the
Galactocentric distribution of SNRs using a sample of bright remnants,
and comparing their distribution in $l$ with than expected from models.
They use several different models, not just a power-low/exponential, and
use a slightly higher surface brightness limit, at 1~GHz, of $1.3 \times
10^{-20}$ $\SigmaUnit$, which gives a somewhat smaller sample of 57
`bright' remnants. \citeauthor{2021MNRAS.504.1536V} prefer a purely
exponential decaying distribution of SNRs with Galactocentric radius.
Their best fit exponential distribution is more concentrated towards the
Galactic centre than the power-law/exponential distribution. The
exponential distribution looks quite different from the
power-law/exponential distribution, with a peak at $R=0$. However, when
considering the integrated surface density, i.e.\ density $\times 2\pi R
\, {\rm d}R$, the exponential result from \citet{2021MNRAS.504.1536V}
and the power-law/exponential from \cite{2015MNRAS.454.1517G} are very
similar (see figure 5 of \citeauthor{2021MNRAS.504.1536V}). It should be
remembered that the volume of the Galaxy close to $R=0$ is small, and
the distribution of SNRs here is not well constrained. This is both due
to small number statistics, and also the fact that the Galactic
background emission is brightest close to $l=0^\circ$. So, near to
$l=0^\circ$ finding SNRs just above the nominal surface brightness
completeness limit is more still difficult, as discussed in
Section~\ref{s:se}. Indeed, even with a slightly higher surface
brightness limit, \citet{2021MNRAS.504.1536V} note their sample of
bright remnants `cannot be assumed to be unbiased' for $|l| < 10^\circ$.

Finally, it should be remembered that the Galactocentric distributions
discussed above are of observed SNRs. This will not be the distribution
of Galactic supernovae if the observability SNRs above the radio surface
brightness limit depends on some property (e.g.\ ISM density) which
varies with Galactocentric radius.

\addcontentsline{toc}{section}{Acknowledgements}
\section*{Acknowledgements}

I am grateful to colleagues for numerous comments on, and corrections to
the various versions of the Galactic SNR catalogue. This research has
made use of the Astrophysics Data System, funded by NASA under
Cooperative Agreement 80NSSC21M00561, the SIMBAD database, CDS,
Strasbourg Astronomical Observatory, France.

%
%

\onecolumn\relax
%
%
\newcount\linesdone
\global\linesdone=0
\newcount\processed
\global\processed=0
%
%
%
\makeatletter
\def\tablefont{\@setfontsize\tablefontsize{10.0}{11.0}}
\makeatother
%
%
\def\captiontext{310 Galactic supernova remnants: summary
  data.\labeltable{t:snrcat}}
\def\tops{%
  \setbox0=\vbox\bgroup\tablefont%
  \centerline{\small{\bf Table~1}. \captiontext}
  \centerline{\hrulefill}
  \vskip-2pt
  \centerline{%
    \hbox to 0.06\hsize{\hfil$l$\enskip}%
    \hbox to 0.065\hsize{\hfil$b$\enskip}%
    \hbox to 0.195\hsize{\hss\quad RA (J2000) Dec\hss}%
    \hbox to 0.12\hsize{\hfil size\hfil}%
    \hbox to 0.05\hsize{type\hfil}%
    \hbox to 0.085\hsize{\quad Flux at\hfil}%
    \hbox to 0.10\hsize{\hfil spectral\hfil}%
    \hbox to 0.329\hsize{\enspace other \hfil}%
    \hfill
  }
  \centerline{%
    \hbox to 0.06\hsize{\hfil/${}^\circ$\kern3pt}%
    \hbox to 0.065\hsize{\hfil/${}^\circ$\kern3pt}%
    \hbox to 0.115\hsize{\hfil/$({\rm h}$\enskip${\rm m}$\enskip${\rm s})$}%
    \hbox to 0.080\hsize{\hfil/$({}^\circ$\kern6pt${}'$)}%
    \hbox to 0.12\hsize{\hfil/arcmin\hfil}%
    \hbox to 0.05\hsize{}%
    \hbox to 0.085\hsize{\ 1~GHz/Jy\hss}%
    \hbox to 0.10\hsize{\hfil index\hfil}%
    \hbox to 0.329\hsize{\enspace name(s)\hfil}%
    \hfill
  }
  \vskip-6pt
  \centerline{\hrulefill} %
  \egroup\box0
  \bgroup
  \tablefont
}
%
%
%
%
%
%
%
\def\LONGITUDE #1 {\def\ldegrees{#1}}
\def\LATITUDE #1 {\def\bdegrees{#1}}
\def\RAHMS #1 #2 #3 {\def\rahms{#1~#2~#3}}
\def\DECDM #1 #2 {\def\decdm{#1~#2}}
\def\SIZE #1 {\edef\size{#1}}            
\def\ALPHA #1 {\def\spectralindex{#1}}
\def\FLUX1GHZ #1 {\def\fluxGHz{#1}}
\def\TYPE #1 {\def\type{#1}}
\def\NAMES #1\par{\def\names{#1}\ifnum\linesdone=0\tops\fi%
  \global\advance\linesdone by 1
  \global\advance\processed by 1
  \centerline{%
    \hbox to 0.06\hsize{\hfil$\ldegrees$}%
    \hbox to 0.065\hsize{\hfil$\bdegrees$}%
    \hbox to 0.11\hsize{\hfil$\rahms$}%
    \hbox to 0.085\hsize{\hfil$\decdm$}%
    \hbox to 0.12\hsize{\FormatSize{\size}\hss}%
    \hbox to 0.05\hsize{\enspace\type\hfil}%
    \hbox to 0.085\hsize{\FormatFlux{\fluxGHz}\hss}%
    \hbox to 0.10\hsize{\quad\spectralindex\hfil}%
    \hbox to 0.325\hsize{\enspace\names\hfil}%
    \hfill
  }
  \vskip -1pt
  \setbox0=\vbox\bgroup\hfuzz=20pt
}
\def\X/{*} 
\let\eightpoint=\footnotesize
%
%
\def\DATE #1\par{
  \egroup
  \ifnum\linesdone=55
    \egroup
    \global\linesdone=0
    \global\processed=0
    \vskip-6pt
    \centerline{\hrulefill}
    \vfill\eject
    \def\captiontext{(continued).}
  \fi
  \ifnum\processed=5
    \vskip 6pt plus 1pt minus 1pt
    \global\processed=0
  \fi
}
\newif\ifQuestion
\newif\ifDecimal
\def\SplitAtDecimal#1.#2\relax{\gdef\BeforeDecimal{#1}\gdef\AfterDecimal{#2}}
\def\EndDecimalTest{}
\def\SplitDecimal#1{\let\next\SplitAtDecimal%
  \expandafter\next#1.\relax%
  \ifx\AfterDecimal\EndDecimalTest%
    \Decimalfalse%
  \else%
    \Decimaltrue%
    \expandafter\next#1\relax%
  \fi}
\def\SplitAtQuestion#1?#2\relax{\gdef\BeforeQuestion{#1}\gdef\AfterQuestion{#2}}
\def\EndQuestionTest{}
\def\SplitQuestion#1{\let\next\SplitAtQuestion%
  \expandafter\next#1?\relax%
  \ifx\AfterQuestion\EndQuestionTest%
    \Questionfalse%
  \else%
    \Questiontrue%
    \expandafter\next#1\relax%
  \fi}
\def\FormatFlux#1{\Decimalfalse\Questionfalse\SplitDecimal{#1}%
   \ifDecimal%
     \setbox0=\hbox{\hbox to 0.045\hsize{\hfill\BeforeDecimal}\hbox
       to 0.040\hsize{.\AfterDecimal\hfill}}%
   \else%
     \SplitQuestion{#1}%
     \ifQuestion%
       \ifx\BeforeQuestion\EndQuestionTest%
         \setbox0=\hbox{\hbox to 0.045\hsize{\hfill?}\hbox
           to 0.040\hsize{\hfill}}%
       \else%
         \setbox0=\hbox{\hbox to 0.045\hsize{\hfill\BeforeQuestion}\hbox
           to 0.040\hsize{?\hfill}}%
        \fi%
     \else%
       \setbox0=\hbox{\hbox to 0.045\hsize{\hfill#1}\hbox
         to 0.040\hsize{\hfill}}%
     \fi%
   \fi\box0}
%
%
\newif\ifStar
\def\SplitAtStar#1*#2\relax{\gdef\BeforeStar{#1}\gdef\AfterStar{#2}}
\def\EndStarTest{}
\def\SplitStar#1{\let\next\SplitAtStar%
  \expandafter\next#1*\relax%
  \ifx\AfterStar\EndStarTest%
    \Starfalse%
  \else%
    \Startrue%
    \expandafter\next#1\relax%
  \fi}
\def\FormatSize#1{\Starfalse\Decimalfalse\Questionfalse\SplitStar{#1}%
   \ifStar%
     \setbox0=\hbox{\hbox
       to 0.065\hsize{\hfill\BeforeStar${\scriptstyle\times}$}\hbox
       to 0.055\hsize{\AfterStar\hfill}}%
   \else%
     \SplitDecimal{#1}%
     \ifDecimal%
       \setbox0=\hbox{\hbox to 0.07\hsize{\hfill\BeforeDecimal}\hbox
         to 0.05\hsize{.\AfterDecimal\hfill}}%
     \else%
       \SplitQuestion{#1}%
       \ifQuestion%
         \ifx\BeforeQuestion\EndQuestionTest%
           \setbox0=\hbox{\hbox to 0.07\hsize{\hfill?}\hbox
             to 0.05\hsize{\hfill}}%
         \else%
           \setbox0=\hbox{\hbox to 0.07\hsize{\hfill\BeforeQuestion}\hbox
             to 0.05\hsize{?\hfill}}%
          \fi%
       \else%
         \setbox0=\hbox{\hbox to 0.07\hsize{\hfill#1}\hbox
           to 0.05\hsize{\hfill}}%
       \fi%
     \fi%
   \fi\box0}
\def\endsnrcat{\egroup\vskip-6pt\centerline{\hrulefill}}

\LONGITUDE 0.0 \LATITUDE +0.0
\RAHMS 17 45 44 \DECDM -29 00
\SIZE 3.5\X/2.5 \TYPE S
\FLUX1GHZ 100? \ALPHA 0.8?
\NAMES Sgr A East

\DATE 20 Dec 2022

\LONGITUDE 0.3 \LATITUDE +0.0
\RAHMS 17 46 15 \DECDM -28 38
\SIZE 15\X/8 \TYPE S
\FLUX1GHZ 22 \ALPHA 0.6
\NAMES

\DATE 8 Jan 2024

\LONGITUDE 0.9 \LATITUDE +0.1
\RAHMS 17 47 21 \DECDM -28 09
\SIZE 8 \TYPE C
\FLUX1GHZ 18? \ALPHA varies
\NAMES

\DATE 5 Jan 2024

\LONGITUDE 1.0 \LATITUDE -0.1
\RAHMS 17 48 30 \DECDM -28 09
\SIZE 8 \TYPE S
\FLUX1GHZ 15 \ALPHA 0.6?
\NAMES

\DATE 20 Dec 2022

\LONGITUDE 1.4 \LATITUDE -0.1
\RAHMS 17 49 39 \DECDM -27 46
\SIZE 10 \TYPE S
\FLUX1GHZ 2? \ALPHA ?
\NAMES

\DATE 20 Dec 2022

\LONGITUDE 1.9 \LATITUDE +0.3
\RAHMS 17 48 45 \DECDM -27 10
\SIZE 1.5 \TYPE S
\FLUX1GHZ 0.6 \ALPHA 0.6
\NAMES

\DATE 12 Mar 2024

\LONGITUDE 3.1 \LATITUDE -0.6
\RAHMS 17 55 30 \DECDM -26 35
\SIZE 52\X/28 \TYPE S
\FLUX1GHZ 5 \ALPHA 0.9?
\NAMES

\DATE 20 Dec 2022

\LONGITUDE 3.7 \LATITUDE -0.2
\RAHMS 17 55 26 \DECDM -25 50
\SIZE 14\X/11 \TYPE S
\FLUX1GHZ 2.3 \ALPHA 0.65
\NAMES

\DATE 20 Dec 2022

\LONGITUDE 3.8 \LATITUDE +0.3
\RAHMS 17 52 55 \DECDM -25 28
\SIZE 18 \TYPE S?
\FLUX1GHZ 3? \ALPHA 0.6
\NAMES

\DATE 20 Dec 2022

\LONGITUDE 4.2 \LATITUDE -3.5
\RAHMS 18 08 55 \DECDM -27 03
\SIZE 28 \TYPE S
\FLUX1GHZ 3.2? \ALPHA 0.6?
\NAMES

\DATE 25 Apr 2014

\LONGITUDE 4.5 \LATITUDE +6.8
\RAHMS 17 30 42 \DECDM -21 29
\SIZE 3 \TYPE S
\FLUX1GHZ 19 \ALPHA 0.64
\NAMES Kepler, SN1604, 3C358

\DATE 5 Jan 2024

\LONGITUDE 4.8 \LATITUDE +6.2
\RAHMS 17 33 25 \DECDM -21 34
\SIZE 18 \TYPE S
\FLUX1GHZ 3 \ALPHA 0.6
\NAMES

\DATE 20 Dec 2023

\LONGITUDE 5.2 \LATITUDE -2.6
\RAHMS 18 07 30 \DECDM -25 45
\SIZE 18 \TYPE S
\FLUX1GHZ 2.6? \ALPHA 0.6?
\NAMES

\DATE 25 Apr 2014

\LONGITUDE 5.4 \LATITUDE -1.2
\RAHMS 18 02 10 \DECDM -24 54
\SIZE 35 \TYPE C?
\FLUX1GHZ 35? \ALPHA 0.2?
\NAMES Milne 56

\DATE 14 Dec 2022

\LONGITUDE 5.5 \LATITUDE +0.3
\RAHMS 17 57 04 \DECDM -24 00
\SIZE 15\X/12 \TYPE S
\FLUX1GHZ 5.5 \ALPHA 0.7
\NAMES

\DATE 11 May 2017

\LONGITUDE 5.9 \LATITUDE +3.1
\RAHMS 17 47 20 \DECDM -22 16
\SIZE 20 \TYPE S
\FLUX1GHZ 3.3? \ALPHA 0.4?
\NAMES

\DATE 9 Dec 2022

\LONGITUDE 6.1 \LATITUDE +0.5
\RAHMS 17 57 29 \DECDM -23 25
\SIZE 18\X/12 \TYPE S
\FLUX1GHZ 4.5 \ALPHA 0.9
\NAMES

\DATE 18 Mar 2024

\LONGITUDE 6.1 \LATITUDE +1.2
\RAHMS 17 54 55 \DECDM -23 05
\SIZE 30\X/26 \TYPE F
\FLUX1GHZ 4.0? \ALPHA 0.3?
\NAMES

\DATE 14 Dec 2022

\LONGITUDE 6.4 \LATITUDE -0.1
\RAHMS 18 00 30 \DECDM -23 26
\SIZE 48 \TYPE C
\FLUX1GHZ 310 \ALPHA varies
\NAMES W28

\DATE 12 Mar 2024

\LONGITUDE 6.4 \LATITUDE +4.0
\RAHMS 17 45 10 \DECDM -21 22
\SIZE 31 \TYPE S
\FLUX1GHZ 1.3? \ALPHA 0.4?
\NAMES

\DATE 25 Apr 2014

\LONGITUDE 6.5 \LATITUDE -0.4
\RAHMS 18 02 11 \DECDM -23 34
\SIZE 18 \TYPE S
\FLUX1GHZ 27 \ALPHA 0.6
\NAMES

\DATE 20 Dec 2022

\LONGITUDE 7.0 \LATITUDE -0.1
\RAHMS 18 01 50 \DECDM -22 54
\SIZE 15 \TYPE S
\FLUX1GHZ 2.5? \ALPHA 0.5?
\NAMES

\DATE 20 Dec 2022

\LONGITUDE 7.2 \LATITUDE +0.2
\RAHMS 18 01 07 \DECDM -22 38
\SIZE 12 \TYPE S
\FLUX1GHZ 2.8 \ALPHA 0.6
\NAMES

\DATE 20 Dec 2022

\LONGITUDE 7.5 \LATITUDE -1.7
\RAHMS 18 10 00 \DECDM -23 10
\SIZE 100 \TYPE S
\FLUX1GHZ 18? \ALPHA 0.7?
\NAMES

\DATE 20 Dec 2022

\LONGITUDE 7.7 \LATITUDE -3.7
\RAHMS 18 17 25 \DECDM -24 04
\SIZE 22 \TYPE S
\FLUX1GHZ 11 \ALPHA 0.32
\NAMES 1814$-$24

\DATE 12 Mar 2024

\LONGITUDE 8.7 \LATITUDE -5.0
\RAHMS 18 24 10 \DECDM -23 48
\SIZE 26 \TYPE S
\FLUX1GHZ 4.4 \ALPHA 0.3
\NAMES

\DATE 13 Dec 2022

\LONGITUDE 8.7 \LATITUDE -0.1
\RAHMS 18 05 30 \DECDM -21 26
\SIZE 45 \TYPE S?
\FLUX1GHZ 80 \ALPHA 0.5
\NAMES (W30)

\DATE 20 Dec 2022

\LONGITUDE 8.9 \LATITUDE +0.4
\RAHMS 18 03 58 \DECDM -21 03
\SIZE 24 \TYPE S
\FLUX1GHZ 9 \ALPHA 0.6
\NAMES

\DATE 20 Dec 2022

\LONGITUDE 9.7 \LATITUDE -0.0
\RAHMS 18 07 22 \DECDM -20 35
\SIZE 15\X/11 \TYPE S
\FLUX1GHZ 3.7 \ALPHA 0.6
\NAMES

\DATE 4 Jan 2024

\LONGITUDE 9.8 \LATITUDE +0.6
\RAHMS 18 05 08 \DECDM -20 14
\SIZE 12 \TYPE S
\FLUX1GHZ 3.9 \ALPHA 0.5
\NAMES

\DATE 20 Dec 2022

\LONGITUDE 9.9 \LATITUDE -0.8
\RAHMS 18 10 41 \DECDM -20 43
\SIZE 12 \TYPE S
\FLUX1GHZ 6.7 \ALPHA 0.4
\NAMES

\DATE 20 Dec 2022

\LONGITUDE 11.0 \LATITUDE -0.0
\RAHMS 18 10 04 \DECDM -19 25
\SIZE 11\X/9 \TYPE S
\FLUX1GHZ 1.3 \ALPHA 0.6
\NAMES

\DATE 20 Dec 2022

\LONGITUDE 11.1 \LATITUDE -0.7
\RAHMS 18 12 46 \DECDM -19 38
\SIZE 11\X/7 \TYPE S
\FLUX1GHZ 1.0 \ALPHA 0.7
\NAMES

\DATE 20 Dec 2022

\LONGITUDE 11.1 \LATITUDE +0.1
\RAHMS 18 09 47 \DECDM -19 12
\SIZE 12\X/10 \TYPE S
\FLUX1GHZ 2.3 \ALPHA 0.4
\NAMES

\DATE 20 Dec 2022

\LONGITUDE 11.2 \LATITUDE -0.3
\RAHMS 18 11 27 \DECDM -19 25
\SIZE 4 \TYPE C
\FLUX1GHZ 22 \ALPHA 0.5
\NAMES

\DATE 12 Mar 2024

\LONGITUDE 11.4 \LATITUDE -0.1
\RAHMS 18 10 47 \DECDM -19 05
\SIZE 8 \TYPE S?
\FLUX1GHZ 6 \ALPHA 0.5
\NAMES

\DATE 20 Dec 2022

\LONGITUDE 11.8 \LATITUDE -0.2
\RAHMS 18 12 25 \DECDM -18 44
\SIZE 4 \TYPE S
\FLUX1GHZ 0.7 \ALPHA 0.3
\NAMES

\DATE 20 Dec 2022

\LONGITUDE 12.0 \LATITUDE -0.1
\RAHMS 18 12 11 \DECDM -18 37
\SIZE 7? \TYPE ?
\FLUX1GHZ 3.5 \ALPHA 0.7
\NAMES

\DATE 20 Dec 2022

\LONGITUDE 12.2 \LATITUDE +0.3
\RAHMS 18 11 17 \DECDM -18 10
\SIZE 6\X/5 \TYPE S
\FLUX1GHZ 0.8 \ALPHA 0.7
\NAMES

\DATE 20 Dec 2022

\LONGITUDE 12.5 \LATITUDE +0.2
\RAHMS 18 12 14 \DECDM -17 55
\SIZE 6\X/5 \TYPE C?
\FLUX1GHZ 0.6 \ALPHA 0.4
\NAMES

\DATE 20 Dec 2022

\LONGITUDE 12.7 \LATITUDE -0.0
\RAHMS 18 13 19 \DECDM -17 54
\SIZE 6 \TYPE S
\FLUX1GHZ 0.8 \ALPHA 0.8
\NAMES

\DATE 20 Dec 2022

\LONGITUDE 12.8 \LATITUDE -0.0
\RAHMS 18 13 37 \DECDM -17 49
\SIZE 3 \TYPE C?
\FLUX1GHZ 0.8 \ALPHA 0.5
\NAMES

\DATE 20 Dec 2022

\LONGITUDE 13.1 \LATITUDE -0.5
\RAHMS 18 16 00 \DECDM -17 49
\SIZE 38\X/28 \TYPE S
\FLUX1GHZ 11? \ALPHA 0.6?
\NAMES

\DATE 20 Dec 2022

\LONGITUDE 13.3 \LATITUDE -1.3
\RAHMS 18 19 20 \DECDM -18 00
\SIZE 70\X/40 \TYPE S?
\FLUX1GHZ ? \ALPHA ?
\NAMES

\DATE 14 Dec 2022

\LONGITUDE 13.5 \LATITUDE +0.2
\RAHMS 18 14 14 \DECDM -17 12
\SIZE 5\X/4 \TYPE S
\FLUX1GHZ 3.5? \ALPHA 1.0?
\NAMES

\DATE 20 Dec 2022

\LONGITUDE 14.1 \LATITUDE -0.1
\RAHMS 18 16 40 \DECDM -16 41
\SIZE 6\X/5 \TYPE S
\FLUX1GHZ 0.5 \ALPHA 0.6
\NAMES

\DATE 20 Dec 2022

\LONGITUDE 15.1 \LATITUDE -1.6
\RAHMS 18 24 00 \DECDM -16 34
\SIZE 30\X/24 \TYPE S?
\FLUX1GHZ 5.5? \ALPHA 0.0?
\NAMES

\DATE 14 Dec 2022

\LONGITUDE 15.4 \LATITUDE +0.1
\RAHMS 18 18 02 \DECDM -15 27
\SIZE 15\X/14 \TYPE C?
\FLUX1GHZ 5.6 \ALPHA 0.62
\NAMES

\DATE 12 Mar 2024

\LONGITUDE 15.5 \LATITUDE -0.1
\RAHMS 18 19 25 \DECDM -15 32
\SIZE 9\X/8 \TYPE ?
\FLUX1GHZ 1.2? \ALPHA 0.55?
\NAMES

\DATE 20 Dec 2022

\LONGITUDE 15.9 \LATITUDE +0.2
\RAHMS 18 18 52 \DECDM -15 02
\SIZE 7\X/5 \TYPE S?
\FLUX1GHZ 5.0 \ALPHA 0.63
\NAMES

\DATE 20 Dec 2022

\LONGITUDE 16.0 \LATITUDE -0.5
\RAHMS 18 21 56 \DECDM -15 14
\SIZE 15\X/10 \TYPE S
\FLUX1GHZ 2.7 \ALPHA 0.6
\NAMES

\DATE 20 Dec 2022

\LONGITUDE 16.2 \LATITUDE -2.7
\RAHMS 18 29 40 \DECDM -16 08
\SIZE 17 \TYPE S
\FLUX1GHZ 2.5 \ALPHA 0.4
\NAMES

\DATE 24 May 2014

\LONGITUDE 16.7 \LATITUDE +0.1
\RAHMS 18 20 56 \DECDM -14 20
\SIZE 4 \TYPE C
\FLUX1GHZ 3.0 \ALPHA 0.6
\NAMES

\DATE 20 Dec 2022

\LONGITUDE 17.0 \LATITUDE -0.0
\RAHMS 18 21 57 \DECDM -14 08
\SIZE 5 \TYPE S
\FLUX1GHZ 0.5 \ALPHA 0.5
\NAMES

\DATE 20 Dec 2022

\LONGITUDE 17.4 \LATITUDE -2.3
\RAHMS 18 30 55 \DECDM -14 52
\SIZE 24? \TYPE S
\FLUX1GHZ 5 \ALPHA 0.5?
\NAMES

\DATE 24 May 2014

\LONGITUDE 17.4 \LATITUDE -0.1
\RAHMS 18 23 08 \DECDM -13 46
\SIZE 6 \TYPE S
\FLUX1GHZ 0.4 \ALPHA 0.7
\NAMES

\DATE 20 Dec 2022

\LONGITUDE 17.8 \LATITUDE -2.6
\RAHMS 18 32 50 \DECDM -14 39
\SIZE 24 \TYPE S
\FLUX1GHZ 5 \ALPHA 0.5
\NAMES

\DATE 24 May 2014

\LONGITUDE 17.8 \LATITUDE +16.7
\RAHMS 17 24 10 \DECDM -05 10
\SIZE 51\X/45 \TYPE S?
\FLUX1GHZ 2.7 \ALPHA 0.8
\NAMES

\DATE 19 Jul 2024

\LONGITUDE 18.1 \LATITUDE -0.1
\RAHMS 18 24 34 \DECDM -13 11
\SIZE 8 \TYPE S
\FLUX1GHZ 4.6 \ALPHA 0.5
\NAMES

\DATE 12 Mar 2024

\LONGITUDE 18.6 \LATITUDE -0.2
\RAHMS 18 25 55 \DECDM -12 50
\SIZE 6 \TYPE S
\FLUX1GHZ 1.4 \ALPHA 0.4
\NAMES

\DATE 20 Dec 2022

\LONGITUDE 18.8 \LATITUDE +0.3
\RAHMS 18 23 58 \DECDM -12 23
\SIZE 17\X/11 \TYPE S
\FLUX1GHZ 33 \ALPHA 0.46
\NAMES Kes 67

\DATE 20 Dec 2022

\LONGITUDE 18.9 \LATITUDE -1.1
\RAHMS 18 29 50 \DECDM -12 58
\SIZE 33 \TYPE C?
\FLUX1GHZ 37 \ALPHA 0.39
\NAMES

\DATE 4 Jan 2024

\LONGITUDE 19.1 \LATITUDE +0.2
\RAHMS 18 24 56 \DECDM -12 07
\SIZE 27 \TYPE S
\FLUX1GHZ 10 \ALPHA 0.5
\NAMES

\DATE 20 Dec 2022

\LONGITUDE 20.0 \LATITUDE -0.2
\RAHMS 18 28 07 \DECDM -11 35
\SIZE 10 \TYPE F
\FLUX1GHZ 10 \ALPHA 0.1
\NAMES

\DATE 20 Dec 2022

\LONGITUDE 21.0 \LATITUDE -0.4
\RAHMS 18 31 12 \DECDM -10 47
\SIZE 9\X/7 \TYPE S
\FLUX1GHZ 1.1 \ALPHA 0.6
\NAMES

\DATE 20 Dec 2022

\LONGITUDE 21.5 \LATITUDE -0.9
\RAHMS 18 33 33 \DECDM -10 35
\SIZE 5 \TYPE C
\FLUX1GHZ 7 \ALPHA varies
\NAMES

\DATE 5 Jan 2024

\LONGITUDE 21.6 \LATITUDE -0.8
\RAHMS 18 33 40 \DECDM -10 25
\SIZE 13 \TYPE S
\FLUX1GHZ 1.4 \ALPHA 0.5?
\NAMES

\DATE 20 Dec 2022

\LONGITUDE 21.8 \LATITUDE -3.0
\RAHMS 18 41 50 \DECDM -11 16
\SIZE 60 \TYPE S
\FLUX1GHZ 5 \ALPHA 0.7
\NAMES

\DATE 20 Dec 2022

\LONGITUDE 21.8 \LATITUDE -0.6
\RAHMS 18 32 45 \DECDM -10 08
\SIZE 20 \TYPE S
\FLUX1GHZ 65 \ALPHA 0.56
\NAMES Kes 69

\DATE 12 Mar 2024

\LONGITUDE 22.7 \LATITUDE -0.2
\RAHMS 18 33 15 \DECDM -09 13
\SIZE 26 \TYPE S?
\FLUX1GHZ 33 \ALPHA 0.6
\NAMES

\DATE 20 Dec 2022

\LONGITUDE 23.3 \LATITUDE -0.3
\RAHMS 18 34 45 \DECDM -08 48
\SIZE 27 \TYPE S
\FLUX1GHZ 70 \ALPHA 0.5
\NAMES W41

\DATE 20 Dec 2022

\LONGITUDE 24.7 \LATITUDE -0.6
\RAHMS 18 38 43 \DECDM -07 32
\SIZE 15? \TYPE S?
\FLUX1GHZ 8 \ALPHA 0.5
\NAMES

\DATE 20 Dec 2022

\LONGITUDE 24.7 \LATITUDE +0.6
\RAHMS 18 34 10 \DECDM -07 05
\SIZE 30\X/15 \TYPE C?
\FLUX1GHZ 20? \ALPHA 0.2?
\NAMES

\DATE 20 Dec 2022

\LONGITUDE 25.1 \LATITUDE -2.3
\RAHMS 18 45 10 \DECDM -08 00
\SIZE 80\X/30? \TYPE S
\FLUX1GHZ 8 \ALPHA 0.5?
\NAMES

\DATE 14 Dec 2022

\LONGITUDE 27.4 \LATITUDE +0.0
\RAHMS 18 41 19 \DECDM -04 56
\SIZE 4 \TYPE S
\FLUX1GHZ 6 \ALPHA 0.68
\NAMES 4C$-$04.71

\DATE 20 Dec 2022

\LONGITUDE 27.8 \LATITUDE +0.6
\RAHMS 18 39 50 \DECDM -04 24
\SIZE 50\X/30 \TYPE F
\FLUX1GHZ 30 \ALPHA varies
\NAMES

\DATE 14 Dec 2022

\LONGITUDE 28.3 \LATITUDE +0.2
\RAHMS 18 42 30 \DECDM -03 58
\SIZE 10 \TYPE S
\FLUX1GHZ 1.3? \ALPHA 0.7?
\NAMES

\DATE 18 Mar 2024

\LONGITUDE 28.6 \LATITUDE -0.1
\RAHMS 18 43 55 \DECDM -03 53
\SIZE 13\X/9 \TYPE S
\FLUX1GHZ 3? \ALPHA ?
\NAMES

\DATE 19 Mar 2024

\LONGITUDE 28.7 \LATITUDE -0.4
\RAHMS 18 45 30 \DECDM -03 54
\SIZE 9 \TYPE S
\FLUX1GHZ 0.9? \ALPHA 0.8?
\NAMES

\DATE 20 Dec 2022

\LONGITUDE 28.8 \LATITUDE +1.5
\RAHMS 18 39 00 \DECDM -02 55
\SIZE 100? \TYPE S?
\FLUX1GHZ ? \ALPHA 0.4?
\NAMES

\DATE 11 May 2017

\LONGITUDE 29.3 \LATITUDE +0.1
\RAHMS 18 44 36 \DECDM -03 06
\SIZE 10? \TYPE C?
\FLUX1GHZ 2.5? \ALPHA ?
\NAMES

\DATE 19 Jul 2024

\LONGITUDE 29.6 \LATITUDE +0.1
\RAHMS 18 44 52 \DECDM -02 57
\SIZE 5 \TYPE S
\FLUX1GHZ 0.5? \ALPHA 0.5?
\NAMES

\DATE 19 Mar 2024

\LONGITUDE 29.7 \LATITUDE -0.3
\RAHMS 18 46 25 \DECDM -02 59
\SIZE 3 \TYPE C
\FLUX1GHZ 9 \ALPHA 0.7
\NAMES Kes 75

\DATE 19 Mar 2024

\LONGITUDE 30.7 \LATITUDE -2.0
\RAHMS 18 54 25 \DECDM -02 54
\SIZE 16 \TYPE ?
\FLUX1GHZ 0.5? \ALPHA 0.7?
\NAMES

\DATE 13 Aug 1998

\LONGITUDE 30.7 \LATITUDE +1.0
\RAHMS 18 44 00 \DECDM -01 32
\SIZE 24\X/18 \TYPE S?
\FLUX1GHZ 6 \ALPHA 0.4
\NAMES

\DATE 14 Dec 2022

\LONGITUDE 31.5 \LATITUDE -0.6
\RAHMS 18 51 10 \DECDM -01 31
\SIZE 18? \TYPE S?
\FLUX1GHZ 2? \ALPHA ?
\NAMES

\DATE 18 Mar 2024

\LONGITUDE 31.9 \LATITUDE +0.0
\RAHMS 18 49 25 \DECDM -00 55
\SIZE 7\X/5 \TYPE S
\FLUX1GHZ 25 \ALPHA varies
\NAMES 3C391

\DATE 18 Mar 2024

\LONGITUDE 32.0 \LATITUDE -4.9
\RAHMS 19 06 00 \DECDM -03 00
\SIZE 60? \TYPE S?
\FLUX1GHZ 22? \ALPHA 0.5?
\NAMES 3C396.1

\DATE 25 Feb 1988

\LONGITUDE 32.1 \LATITUDE -0.9
\RAHMS 18 53 10 \DECDM -01 08
\SIZE 40? \TYPE C?
\FLUX1GHZ 4? \ALPHA 0.7?
\NAMES

\DATE 19 Mar 2024

\LONGITUDE 32.4 \LATITUDE +0.1
\RAHMS 18 50 05 \DECDM -00 25
\SIZE 6 \TYPE S
\FLUX1GHZ 0.8? \ALPHA 0.2?
\NAMES

\DATE 19 Mar 2024

\LONGITUDE 32.8 \LATITUDE -0.1
\RAHMS 18 51 25 \DECDM -00 08
\SIZE 22\X/15 \TYPE S?
\FLUX1GHZ 12 \ALPHA 0.3
\NAMES Kes 78

\DATE 19 Mar 2024

\LONGITUDE 33.2 \LATITUDE -0.6
\RAHMS 18 53 50 \DECDM -00 02
\SIZE 18 \TYPE S
\FLUX1GHZ 3 \ALPHA 0.3
\NAMES

\DATE 19 Mar 2024

\LONGITUDE 33.6 \LATITUDE +0.1
\RAHMS 18 52 48 \DECDM +00 41
\SIZE 10 \TYPE S
\FLUX1GHZ 20 \ALPHA 0.51
\NAMES Kes 79, 4C00.70, HC13

\DATE 19 Mar 2024

\LONGITUDE 34.7 \LATITUDE -0.4
\RAHMS 18 56 00 \DECDM +01 22
\SIZE 35\X/27 \TYPE C
\FLUX1GHZ 240 \ALPHA 0.37
\NAMES W44, 3C392

\DATE 19 Mar 2024

\LONGITUDE 35.6 \LATITUDE -0.4
\RAHMS 18 57 55 \DECDM +02 13
\SIZE 15\X/11 \TYPE S?
\FLUX1GHZ 9 \ALPHA varies
\NAMES

\DATE 19 Mar 2024

\LONGITUDE 36.6 \LATITUDE -0.7
\RAHMS 19 00 35 \DECDM +02 56
\SIZE 25? \TYPE S?
\FLUX1GHZ 1.0 \ALPHA 0.7?
\NAMES

\DATE 20 Dec 2022

\LONGITUDE 36.6 \LATITUDE +2.6
\RAHMS 18 48 49 \DECDM +04 26
\SIZE 17\X/13? \TYPE S
\FLUX1GHZ 0.7? \ALPHA 0.5?
\NAMES

\DATE 19 May 1992

\LONGITUDE 38.7 \LATITUDE -1.3
\RAHMS 19 06 40 \DECDM +04 28
\SIZE 32\X/19? \TYPE S
\FLUX1GHZ ? \ALPHA ?
\NAMES

\DATE 14 Dec 2022

\LONGITUDE 39.2 \LATITUDE -0.3
\RAHMS 19 04 08 \DECDM +05 28
\SIZE 8\X/6 \TYPE C
\FLUX1GHZ 18 \ALPHA 0.34
\NAMES 3C396, HC24, NRAO 593

\DATE 5 Jan 2024

\LONGITUDE 39.7 \LATITUDE -2.0
\RAHMS 19 12 20 \DECDM +04 55
\SIZE 120\X/60 \TYPE ?
\FLUX1GHZ 85? \ALPHA 0.7?
\NAMES W50, SS433

\DATE 4 Jan 2024

\LONGITUDE 40.5 \LATITUDE -0.5
\RAHMS 19 07 10 \DECDM +06 31
\SIZE 22 \TYPE S
\FLUX1GHZ 11 \ALPHA 0.4
\NAMES

\DATE 20 Dec 2022

\LONGITUDE 41.1 \LATITUDE -0.3
\RAHMS 19 07 34 \DECDM +07 08
\SIZE 4.5\X/2.5 \TYPE S
\FLUX1GHZ 25 \ALPHA 0.50
\NAMES 3C397

\DATE 5 Jan 2024

\LONGITUDE 41.5 \LATITUDE +0.4
\RAHMS 19 05 50 \DECDM +07 46
\SIZE 10 \TYPE S?
\FLUX1GHZ 1? \ALPHA ?
\NAMES

\DATE 20 Dec 2022

\LONGITUDE 42.0 \LATITUDE -0.1
\RAHMS 19 08 10 \DECDM +08 00
\SIZE 8 \TYPE S?
\FLUX1GHZ 0.5? \ALPHA ?
\NAMES

\DATE 20 Dec 2022

\LONGITUDE 42.8 \LATITUDE +0.6
\RAHMS 19 07 20 \DECDM +09 05
\SIZE 24 \TYPE S
\FLUX1GHZ 3? \ALPHA 0.5?
\NAMES

\DATE 20 Dec 2022

\LONGITUDE 43.3 \LATITUDE -0.2
\RAHMS 19 11 08 \DECDM +09 06
\SIZE 4\X/3 \TYPE S
\FLUX1GHZ 38 \ALPHA 0.46
\NAMES W49B

\DATE 5 Jan 2024

\LONGITUDE 43.9 \LATITUDE +1.6
\RAHMS 19 05 50 \DECDM +10 30
\SIZE 60? \TYPE S?
\FLUX1GHZ 9.0 \ALPHA 0.5
\NAMES

\DATE 14 Dec 2022

\LONGITUDE 45.7 \LATITUDE -0.4
\RAHMS 19 16 25 \DECDM +11 09
\SIZE 22 \TYPE S
\FLUX1GHZ 4.2? \ALPHA 0.4?
\NAMES

\DATE 20 Dec 2022

\LONGITUDE 46.8 \LATITUDE -0.3
\RAHMS 19 18 10 \DECDM +12 09
\SIZE 15 \TYPE S
\FLUX1GHZ 16 \ALPHA 0.54
\NAMES (HC30)

\DATE 12 Mar 2024

\LONGITUDE 49.2 \LATITUDE -0.7
\RAHMS 19 23 50 \DECDM +14 06
\SIZE 30 \TYPE S?
\FLUX1GHZ 160? \ALPHA 0.3?
\NAMES (W51)

\DATE 12 Mar 2024

\LONGITUDE 53.4 \LATITUDE +0.0
\RAHMS 19 29 57 \DECDM +18 10
\SIZE 10? \TYPE S
\FLUX1GHZ 1.5 \ALPHA 0.6?
\NAMES

\DATE 4 Jan 2024

\LONGITUDE 53.6 \LATITUDE -2.2
\RAHMS 19 38 50 \DECDM +17 14
\SIZE 33\X/28 \TYPE S
\FLUX1GHZ 8 \ALPHA 0.50
\NAMES 3C400.2, NRAO 611

\DATE 16 Dec 2022

\LONGITUDE 54.1 \LATITUDE +0.3
\RAHMS 19 30 31 \DECDM +18 52
\SIZE 12? \TYPE C?
\FLUX1GHZ 0.5 \ALPHA 0.1
\NAMES

\DATE 12 Mar 2024

\LONGITUDE 54.4 \LATITUDE -0.3
\RAHMS 19 33 20 \DECDM +18 56
\SIZE 40 \TYPE S
\FLUX1GHZ 28 \ALPHA 0.5
\NAMES (HC40)

\DATE 20 Dec 2022

\LONGITUDE 55.0 \LATITUDE +0.3
\RAHMS 19 32 00 \DECDM +19 50
\SIZE 20\X/15? \TYPE S
\FLUX1GHZ 0.5? \ALPHA 0.5?
\NAMES

\DATE 20 Dec 2022

\LONGITUDE 55.7 \LATITUDE +3.4
\RAHMS 19 21 20 \DECDM +21 44
\SIZE 23 \TYPE S
\FLUX1GHZ 1? \ALPHA 0.3?
\NAMES

\DATE 24 May 2014

\LONGITUDE 57.2 \LATITUDE +0.8
\RAHMS 19 34 59 \DECDM +21 57
\SIZE 12? \TYPE S?
\FLUX1GHZ 1.8 \ALPHA 0.35
\NAMES (4C21.53)

\DATE 20 Dec 2022

\LONGITUDE 59.5 \LATITUDE +0.1
\RAHMS 19 42 33 \DECDM +23 35
\SIZE 15 \TYPE S
\FLUX1GHZ 3? \ALPHA ?
\NAMES

\DATE 20 Dec 2022

\LONGITUDE 63.7 \LATITUDE +1.1
\RAHMS 19 47 52 \DECDM +27 45
\SIZE 8 \TYPE F
\FLUX1GHZ 1.8 \ALPHA 0.24
\NAMES

\DATE 9 Apr 2019

\LONGITUDE 64.5 \LATITUDE +0.9
\RAHMS 19 50 25 \DECDM +28 16
\SIZE 8 \TYPE S?
\FLUX1GHZ 0.15? \ALPHA 0.5
\NAMES

\DATE 5 Aug 2022

\LONGITUDE 65.1 \LATITUDE +0.6
\RAHMS 19 54 40 \DECDM +28 35
\SIZE 90\X/50 \TYPE S
\FLUX1GHZ 5.5 \ALPHA 0.61
\NAMES

\DATE 5 Jan 2024

\LONGITUDE 65.3 \LATITUDE +5.7
\RAHMS 19 33 00 \DECDM +31 10
\SIZE 310\X/240 \TYPE S?
\FLUX1GHZ 42 \ALPHA 0.6
\NAMES

\DATE 13 Dec 2022

\LONGITUDE 65.7 \LATITUDE +1.2
\RAHMS 19 52 10 \DECDM +29 26
\SIZE 22 \TYPE F
\FLUX1GHZ 5.1 \ALPHA varies
\NAMES DA 495

\DATE 9 Dec 2022

\LONGITUDE 66.0 \LATITUDE -0.0
\RAHMS 19 57 50 \DECDM +29 03
\SIZE 31\X/25? \TYPE S
\FLUX1GHZ ? \ALPHA ?
\NAMES

\DATE 14 Dec 2022

\LONGITUDE 67.6 \LATITUDE +0.9
\RAHMS 19 57 45 \DECDM +30 53
\SIZE 50\X/45? \TYPE S
\FLUX1GHZ ? \ALPHA ?
\NAMES

\DATE 9 Apr 2019

\LONGITUDE 67.7 \LATITUDE +1.8
\RAHMS 19 54 32 \DECDM +31 29
\SIZE 15\X/12 \TYPE S
\FLUX1GHZ 1.0 \ALPHA 0.61
\NAMES

\DATE 9 Apr 2019

\LONGITUDE 67.8 \LATITUDE +0.5
\RAHMS 20 00 00 \DECDM +30 51
\SIZE 7\X/5 \TYPE ?
\FLUX1GHZ ? \ALPHA ?
\NAMES

\DATE 13 Dec 2022

\LONGITUDE 68.6 \LATITUDE -1.2
\RAHMS 20 08 40 \DECDM +30 37
\SIZE 23 \TYPE ?
\FLUX1GHZ 1.1 \ALPHA 0.2
\NAMES

\DATE 24 May 2014

\LONGITUDE 69.0 \LATITUDE +2.7
\RAHMS 19 53 20 \DECDM +32 55
\SIZE 80? \TYPE ?
\FLUX1GHZ 120? \ALPHA varies
\NAMES CTB 80

\DATE 16 Dec 2022

\LONGITUDE 69.7 \LATITUDE +1.0
\RAHMS 20 02 40 \DECDM +32 43
\SIZE 16\X/14 \TYPE S
\FLUX1GHZ 2.0 \ALPHA 0.7
\NAMES

\DATE 28 Apr 2014

\LONGITUDE 70.0 \LATITUDE -21.5
\RAHMS 21 24 00 \DECDM +19 23
\SIZE 330\X/240 \TYPE S
\FLUX1GHZ ? \ALPHA ?
\NAMES

\DATE 13 Dec 2022

\LONGITUDE 73.9 \LATITUDE +0.9
\RAHMS 20 14 15 \DECDM +36 12
\SIZE 27 \TYPE S?
\FLUX1GHZ 9 \ALPHA 0.23
\NAMES

\DATE 14 Dec 2022

\LONGITUDE 74.0 \LATITUDE -8.5
\RAHMS 20 51 00 \DECDM +30 40
\SIZE 230\X/160 \TYPE S
\FLUX1GHZ 210 \ALPHA varies
\NAMES Cygnus Loop

\DATE 12 Mar 2024

\LONGITUDE 74.9 \LATITUDE +1.2
\RAHMS 20 16 02 \DECDM +37 12
\SIZE 8\X/6 \TYPE F
\FLUX1GHZ 9 \ALPHA 0.3
\NAMES CTB 87

\DATE 4 Jan 2024

\LONGITUDE 76.9 \LATITUDE +1.0
\RAHMS 20 22 20 \DECDM +38 43
\SIZE 9 \TYPE C
\FLUX1GHZ 2? \ALPHA ?
\NAMES

\DATE 9 Apr 2019

\LONGITUDE 78.2 \LATITUDE +2.1
\RAHMS 20 20 50 \DECDM +40 26
\SIZE 60 \TYPE S
\FLUX1GHZ 320 \ALPHA 0.51
\NAMES DR4, $\gamma$ Cygni SNR

\DATE 12 Mar 2024

\LONGITUDE 82.2 \LATITUDE +5.3
\RAHMS 20 19 00 \DECDM +45 30
\SIZE 95\X/65 \TYPE S
\FLUX1GHZ 120? \ALPHA 0.5?
\NAMES W63

\DATE 13 Dec 2022

\LONGITUDE 83.0 \LATITUDE -0.3
\RAHMS 20 46 55 \DECDM +42 52
\SIZE 9\X/7 \TYPE S
\FLUX1GHZ 1 \ALPHA 0.4
\NAMES

\DATE 9 Apr 2019

\LONGITUDE 84.2 \LATITUDE -0.8
\RAHMS 20 53 20 \DECDM +43 27
\SIZE 20\X/16 \TYPE S
\FLUX1GHZ 11 \ALPHA 0.5
\NAMES

\DATE 28 Apr 2014

\LONGITUDE 85.4 \LATITUDE +0.7
\RAHMS 20 50 40 \DECDM +45 22
\SIZE 24? \TYPE S
\FLUX1GHZ ? \ALPHA 0.2
\NAMES

\DATE 14 Dec 2022

\LONGITUDE 85.9 \LATITUDE -0.6
\RAHMS 20 58 40 \DECDM +44 53
\SIZE 24 \TYPE S
\FLUX1GHZ ? \ALPHA 0.2
\NAMES

\DATE 14 Dec 2022

\LONGITUDE 89.0 \LATITUDE +4.7
\RAHMS 20 45 00 \DECDM +50 35
\SIZE 120\X/90 \TYPE S
\FLUX1GHZ 220 \ALPHA 0.38
\NAMES HB21

\DATE 16 Dec 2022

\LONGITUDE 93.3 \LATITUDE +6.9
\RAHMS 20 52 25 \DECDM +55 21
\SIZE 27\X/20 \TYPE C?
\FLUX1GHZ 9 \ALPHA 0.45
\NAMES DA 530, 4C(T)55.38.1

\DATE 12 Mar 2024

\LONGITUDE 93.7 \LATITUDE -0.2
\RAHMS 21 29 20 \DECDM +50 50
\SIZE 80 \TYPE S
\FLUX1GHZ 65 \ALPHA 0.65
\NAMES CTB 104A, DA 551

\DATE 14 Dec 2022

\LONGITUDE 94.0 \LATITUDE +1.0
\RAHMS 21 24 50 \DECDM +51 53
\SIZE 30\X/25 \TYPE S
\FLUX1GHZ 13 \ALPHA 0.45
\NAMES 3C434.1

\DATE 13 Dec 2022

\LONGITUDE 96.0 \LATITUDE +2.0
\RAHMS 21 30 30 \DECDM +53 59
\SIZE 26 \TYPE S
\FLUX1GHZ 0.35 \ALPHA 0.6
\NAMES

\DATE 24 May 2014

\LONGITUDE 106.3 \LATITUDE +2.7
\RAHMS 22 27 30 \DECDM +60 50
\SIZE 60\X/24 \TYPE C?
\FLUX1GHZ 6 \ALPHA 0.6
\NAMES

\DATE 12 Mar 2024

\LONGITUDE 107.0 \LATITUDE +9.0
\RAHMS 22 01 00 \DECDM +66 30
\SIZE 180? \TYPE ?
\FLUX1GHZ 11? \ALPHA 0.9?
\NAMES

\DATE 20 Dec 2022

\LONGITUDE 108.2 \LATITUDE -0.6
\RAHMS 22 53 40 \DECDM +58 50
\SIZE 70\X/54 \TYPE S
\FLUX1GHZ 8 \ALPHA 0.5
\NAMES

\DATE 13 Dec 2022

\LONGITUDE 109.1 \LATITUDE -1.0
\RAHMS 23 01 35 \DECDM +58 53
\SIZE 28 \TYPE S
\FLUX1GHZ 20 \ALPHA 0.45
\NAMES CTB 109

\DATE 15 Dec 2022

\LONGITUDE 111.7 \LATITUDE -2.1
\RAHMS 23 23 26 \DECDM +58 48
\SIZE 5 \TYPE S
\FLUX1GHZ 2300 \ALPHA 0.77
\NAMES Cassiopeia A, 3C461

\DATE 12 Mar 2024

\LONGITUDE 113.0 \LATITUDE +0.2
\RAHMS 23 26 50 \DECDM +61 26
\SIZE 40\X/17? \TYPE ?
\FLUX1GHZ 4 \ALPHA 0.5?
\NAMES

\DATE 28 Apr 2017

\LONGITUDE 114.3 \LATITUDE +0.3
\RAHMS 23 37 00 \DECDM +61 55
\SIZE 90\X/55 \TYPE S
\FLUX1GHZ 5.5 \ALPHA 0.5
\NAMES

\DATE 28 Apr 2014

\LONGITUDE 116.5 \LATITUDE +1.1
\RAHMS 23 53 40 \DECDM +63 15
\SIZE 80\X/60 \TYPE S
\FLUX1GHZ 10 \ALPHA 0.5
\NAMES

\DATE 28 Apr 2014

\LONGITUDE 116.6 \LATITUDE -26.1
\RAHMS 00 23 00 \DECDM +36 30
\SIZE 235 \TYPE S
\FLUX1GHZ ? \ALPHA ?
\NAMES

\DATE 19 Jul 2024

\LONGITUDE 116.9 \LATITUDE +0.2
\RAHMS 23 59 10 \DECDM +62 26
\SIZE 34 \TYPE S
\FLUX1GHZ 8 \ALPHA 0.57
\NAMES CTB 1

\DATE 12 Mar 2024

\LONGITUDE 119.5 \LATITUDE +10.2
\RAHMS 00 06 40 \DECDM +72 45
\SIZE 90? \TYPE S
\FLUX1GHZ 36 \ALPHA 0.6
\NAMES CTA 1

\DATE 9 Apr 2019

\LONGITUDE 120.1 \LATITUDE +1.4
\RAHMS 00 25 18 \DECDM +64 09
\SIZE 8 \TYPE S
\FLUX1GHZ 50 \ALPHA 0.58
\NAMES Tycho, 3C10, SN1572

\DATE 12 Mar 2024

\LONGITUDE 126.2 \LATITUDE +1.6
\RAHMS 01 22 00 \DECDM +64 15
\SIZE 70 \TYPE S?
\FLUX1GHZ 6 \ALPHA 0.5
\NAMES

\DATE 12 Mar 2024

\LONGITUDE 127.1 \LATITUDE +0.5
\RAHMS 01 28 20 \DECDM +63 10
\SIZE 45 \TYPE S
\FLUX1GHZ 12 \ALPHA 0.45
\NAMES R5

\DATE 12 Mar 2024

\LONGITUDE 130.7 \LATITUDE +3.1
\RAHMS 02 05 41 \DECDM +64 49
\SIZE 9\X/5 \TYPE F
\FLUX1GHZ 33 \ALPHA 0.07
\NAMES 3C58, SN1181

\DATE 15 Dec 2022

\LONGITUDE 132.7 \LATITUDE +1.3
\RAHMS 02 17 40 \DECDM +62 45
\SIZE 80 \TYPE S
\FLUX1GHZ 45 \ALPHA 0.6
\NAMES HB3

\DATE 4 Jan 2024

\LONGITUDE 150.3 \LATITUDE +4.5
\RAHMS 04 27 00 \DECDM +55 28
\SIZE 180\X/150 \TYPE S
\FLUX1GHZ ? \ALPHA ?
\NAMES

\DATE 18 Mar 2024

\LONGITUDE 152.4 \LATITUDE -2.1
\RAHMS 04 07 50 \DECDM +49 11
\SIZE 100\X/95 \TYPE S
\FLUX1GHZ 3.5? \ALPHA 0.7?
\NAMES

\DATE 18 Mar 2024

\LONGITUDE 156.2 \LATITUDE +5.7
\RAHMS 04 58 40 \DECDM +51 50
\SIZE 110 \TYPE S
\FLUX1GHZ 5 \ALPHA 0.5
\NAMES

\DATE 18 Mar 2024

\LONGITUDE 159.6 \LATITUDE +7.3
\RAHMS 05 20 00 \DECDM +50 00
\SIZE 240\X/180? \TYPE S
\FLUX1GHZ ? \ALPHA ?
\NAMES

\DATE 8 Jun 2019

\LONGITUDE 160.9 \LATITUDE +2.6
\RAHMS 05 01 00 \DECDM +46 40
\SIZE 140\X/120 \TYPE S
\FLUX1GHZ 110 \ALPHA 0.64
\NAMES HB9

\DATE 18 Mar 2024

\LONGITUDE 166.0 \LATITUDE +4.3
\RAHMS 05 26 30 \DECDM +42 56
\SIZE 55\X/35 \TYPE S
\FLUX1GHZ 7 \ALPHA 0.37
\NAMES VRO 42.05.01

\DATE 12 Mar 2024

\LONGITUDE 178.2 \LATITUDE -4.2
\RAHMS 05 25 05 \DECDM +28 11
\SIZE 72\X/62 \TYPE S
\FLUX1GHZ 2 \ALPHA 0.5
\NAMES

\DATE 28 Apr 2017

\LONGITUDE 179.0 \LATITUDE +2.6
\RAHMS 05 53 40 \DECDM +31 05
\SIZE 70 \TYPE S?
\FLUX1GHZ 7 \ALPHA 0.4
\NAMES

\DATE 8 Jun 2019

\LONGITUDE 180.0 \LATITUDE -1.7
\RAHMS 05 39 00 \DECDM +27 50
\SIZE 180 \TYPE S
\FLUX1GHZ 65 \ALPHA varies
\NAMES S147

\DATE 4 Jan 2024

\LONGITUDE 181.1 \LATITUDE +9.5
\RAHMS 06 26 40 \DECDM +32 30
\SIZE 74 \TYPE S
\FLUX1GHZ 0.4? \ALPHA 0.4?
\NAMES

\DATE 5 Aug 2022

\LONGITUDE 182.4 \LATITUDE +4.3
\RAHMS 06 08 10 \DECDM +29 00
\SIZE 50 \TYPE S
\FLUX1GHZ 0.5 \ALPHA 0.4
\NAMES

\DATE 13 Dec 2022

\LONGITUDE 184.6 \LATITUDE -5.8
\RAHMS 05 34 31 \DECDM +22 01
\SIZE 7\X/5 \TYPE F
\FLUX1GHZ 900 \ALPHA 0.30
\NAMES Crab Nebula, 3C144, SN1054

\DATE 12 Mar 2024

\LONGITUDE 189.1 \LATITUDE +3.0
\RAHMS 06 17 00 \DECDM +22 34
\SIZE 45 \TYPE C
\FLUX1GHZ 165 \ALPHA 0.36
\NAMES IC443, 3C157

\DATE 12 Mar 2024

\LONGITUDE 189.6 \LATITUDE +3.3
\RAHMS 06 19 40 \DECDM +22 00
\SIZE 90? \TYPE S
\FLUX1GHZ ? \ALPHA ?
\NAMES

\DATE 19 Jul 2024

\LONGITUDE 190.9 \LATITUDE -2.2
\RAHMS 06 01 55 \DECDM +18 24
\SIZE 70\X/60 \TYPE S
\FLUX1GHZ 1.3? \ALPHA 0.7?
\NAMES

\DATE 13 Dec 2022

\LONGITUDE 203.1 \LATITUDE +6.6
\RAHMS 06 57 00 \DECDM +11 40
\SIZE 150 \TYPE S
\FLUX1GHZ ? \ALPHA ?
\NAMES

\DATE 19 Jul 2024

\LONGITUDE 205.5 \LATITUDE +0.5
\RAHMS 06 39 00 \DECDM +06 30
\SIZE 220 \TYPE S
\FLUX1GHZ 140 \ALPHA 0.4
\NAMES Monoceros Nebula

\DATE 15 Dec 2022

\LONGITUDE 206.7 \LATITUDE +5.9
\RAHMS 07 01 00 \DECDM +08 10
\SIZE 210 \TYPE S
\FLUX1GHZ ? \ALPHA ?
\NAMES

\DATE 19 Jul 2024

\LONGITUDE 206.9 \LATITUDE +2.3
\RAHMS 06 48 40 \DECDM +06 26
\SIZE 60\X/40 \TYPE S?
\FLUX1GHZ 6 \ALPHA 0.5
\NAMES PKS 0646$+$06

\DATE 13 Dec 2022

\LONGITUDE 213.0 \LATITUDE -0.6
\RAHMS 06 50 50 \DECDM -00 30
\SIZE 160\X/140? \TYPE S
\FLUX1GHZ 21 \ALPHA 0.4
\NAMES

\DATE 15 Dec 2022

\LONGITUDE 249.5 \LATITUDE +24.5
\RAHMS 09 34 00 \DECDM -17 00
\SIZE 260 \TYPE S
\FLUX1GHZ 27 \ALPHA 0.7
\NAMES Hoinga

\DATE 20 Dec 2022

\LONGITUDE 260.4 \LATITUDE -3.4
\RAHMS 08 22 10 \DECDM -43 00
\SIZE 60\X/50 \TYPE S
\FLUX1GHZ 130 \ALPHA 0.5
\NAMES Puppis A, MSH 08$-$4{\sl 4}

\DATE 12 Mar 2024

\LONGITUDE 261.9 \LATITUDE +5.5
\RAHMS 09 04 20 \DECDM -38 42
\SIZE 40\X/30 \TYPE S
\FLUX1GHZ 10? \ALPHA 0.4?
\NAMES

\DATE 13 Aug 1998

\LONGITUDE 263.9 \LATITUDE -3.3
\RAHMS 08 34 00 \DECDM -45 50
\SIZE 255 \TYPE C
\FLUX1GHZ 1750 \ALPHA varies
\NAMES Vela (XYZ)

\DATE 4 Jan 2024

\LONGITUDE 266.2 \LATITUDE -1.2
\RAHMS 08 52 00 \DECDM -46 20
\SIZE 120 \TYPE S
\FLUX1GHZ 50? \ALPHA 0.3?
\NAMES RX J0852.0$-$4622

\DATE 12 Mar 2024

\LONGITUDE 272.2 \LATITUDE -3.2
\RAHMS 09 06 50 \DECDM -52 07
\SIZE 15? \TYPE S?
\FLUX1GHZ 0.4 \ALPHA 0.6
\NAMES

\DATE 12 Mar 2024

\LONGITUDE 279.0 \LATITUDE +1.1
\RAHMS 09 57 40 \DECDM -53 15
\SIZE 95 \TYPE S
\FLUX1GHZ 30? \ALPHA 0.6?
\NAMES

\DATE 19 Dec 2022

\LONGITUDE 284.3 \LATITUDE -1.8
\RAHMS 10 18 15 \DECDM -59 00
\SIZE 24? \TYPE S
\FLUX1GHZ 11? \ALPHA 0.3?
\NAMES MSH 10$-$5{\sl 3}

\DATE 9 Dec 2022

\LONGITUDE 286.5 \LATITUDE -1.2
\RAHMS 10 35 40 \DECDM -59 42
\SIZE 26\X/6 \TYPE S?
\FLUX1GHZ 1.4? \ALPHA ?
\NAMES

\DATE 25 Apr 2014

\LONGITUDE 288.8 \LATITUDE -6.3
\RAHMS 10 30 20 \DECDM -65 15
\SIZE 108\X/96 \TYPE S
\FLUX1GHZ 11 \ALPHA 0.41
\NAMES

\DATE 19 Jul 2024

\LONGITUDE 289.7 \LATITUDE -0.3
\RAHMS 11 01 15 \DECDM -60 18
\SIZE 18\X/14 \TYPE S
\FLUX1GHZ 6.2 \ALPHA 0.2?
\NAMES

\DATE 21 Aug 1996

\LONGITUDE 290.1 \LATITUDE -0.8
\RAHMS 11 03 05 \DECDM -60 56
\SIZE 19\X/14 \TYPE S
\FLUX1GHZ 42 \ALPHA 0.4
\NAMES MSH 11$-$6{\sl 1}A

\DATE 13 May 2017

\LONGITUDE 291.0 \LATITUDE -0.1
\RAHMS 11 11 54 \DECDM -60 38
\SIZE 15\X/13 \TYPE C
\FLUX1GHZ 16 \ALPHA 0.29
\NAMES (MSH 11$-$6{\sl 2})

\DATE 13 May 2017

\LONGITUDE 292.0 \LATITUDE +1.8
\RAHMS 11 24 36 \DECDM -59 16
\SIZE 12\X/8 \TYPE C
\FLUX1GHZ 15 \ALPHA 0.4
\NAMES MSH 11$-$5{\sl 4}

\DATE 5 Jan 2024

\LONGITUDE 292.2 \LATITUDE -0.5
\RAHMS 11 19 20 \DECDM -61 28
\SIZE 20\X/15 \TYPE S
\FLUX1GHZ 7 \ALPHA 0.5
\NAMES

\DATE 13 May 2017

\LONGITUDE 293.8 \LATITUDE +0.6
\RAHMS 11 35 00 \DECDM -60 54
\SIZE 20 \TYPE C
\FLUX1GHZ 5? \ALPHA 0.6?
\NAMES

\DATE 21 Aug 1996

\LONGITUDE 294.1 \LATITUDE -0.0
\RAHMS 11 36 10 \DECDM -61 38
\SIZE 40 \TYPE S
\FLUX1GHZ $>$2? \ALPHA ?
\NAMES

\DATE 21 Aug 1996

\LONGITUDE 296.1 \LATITUDE -0.5
\RAHMS 11 51 10 \DECDM -62 34
\SIZE 37\X/25 \TYPE S
\FLUX1GHZ 8? \ALPHA 0.6?
\NAMES

\DATE 5 Jan 2024

\LONGITUDE 296.5 \LATITUDE +10.0
\RAHMS 12 09 40 \DECDM -52 25
\SIZE 90\X/65 \TYPE S
\FLUX1GHZ 48 \ALPHA 0.5
\NAMES PKS 1209$-$51/52

\DATE 4 Jan 2024

\LONGITUDE 296.7 \LATITUDE -0.9
\RAHMS 11 55 30 \DECDM -63 08
\SIZE 15\X/8 \TYPE S
\FLUX1GHZ 3 \ALPHA 0.5
\NAMES

\DATE 12 Jun 2017

\LONGITUDE 296.8 \LATITUDE -0.3
\RAHMS 11 58 30 \DECDM -62 35
\SIZE 20\X/14 \TYPE S
\FLUX1GHZ 9 \ALPHA 0.6
\NAMES 1156$-$62

\DATE 12 Mar 2024

\LONGITUDE 298.5 \LATITUDE -0.3
\RAHMS 12 12 40 \DECDM -62 52
\SIZE 5? \TYPE ?
\FLUX1GHZ 5? \ALPHA 0.4?
\NAMES

\DATE 16 Mar 2009

\LONGITUDE 298.6 \LATITUDE -0.0
\RAHMS 12 13 41 \DECDM -62 37
\SIZE 12\X/9 \TYPE S
\FLUX1GHZ 5? \ALPHA 0.3
\NAMES

\DATE 13 Dec 2022

\LONGITUDE 299.2 \LATITUDE -2.9
\RAHMS 12 15 13 \DECDM -65 30
\SIZE 18\X/11 \TYPE S
\FLUX1GHZ 0.5? \ALPHA ?
\NAMES

\DATE 9 Dec 2022

\LONGITUDE 299.6 \LATITUDE -0.5
\RAHMS 12 21 45 \DECDM -63 09
\SIZE 13 \TYPE S
\FLUX1GHZ 1.0? \ALPHA ?
\NAMES

\DATE 21 Aug 1996

\LONGITUDE 301.4 \LATITUDE -1.0
\RAHMS 12 37 55 \DECDM -63 49
\SIZE 37\X/23 \TYPE S
\FLUX1GHZ 2.1? \ALPHA ?
\NAMES

\DATE 14 Dec 2022

\LONGITUDE 302.3 \LATITUDE +0.7
\RAHMS 12 45 55 \DECDM -62 08
\SIZE 17 \TYPE S
\FLUX1GHZ 5? \ALPHA 0.4?
\NAMES

\DATE 21 Aug 1996

\LONGITUDE 304.6 \LATITUDE +0.1
\RAHMS 13 05 59 \DECDM -62 42
\SIZE 8 \TYPE S
\FLUX1GHZ 14 \ALPHA 0.5
\NAMES Kes 17

\DATE 12 Mar 2024

\LONGITUDE 306.3 \LATITUDE -0.9
\RAHMS 13 21 50 \DECDM -63 34
\SIZE 4 \TYPE S?
\FLUX1GHZ 0.16? \ALPHA 0.5?
\NAMES

\DATE 5 Jan 2024

\LONGITUDE 308.1 \LATITUDE -0.7
\RAHMS 13 37 37 \DECDM -63 04
\SIZE 13 \TYPE S
\FLUX1GHZ 1.2? \ALPHA ?
\NAMES

\DATE 21 Aug 1996

\LONGITUDE 308.4 \LATITUDE -1.4
\RAHMS 13 41 30 \DECDM -63 44
\SIZE 12\X/6? \TYPE S?
\FLUX1GHZ 0.4? \ALPHA ?
\NAMES

\DATE 16 Dec 2022

\LONGITUDE 308.8 \LATITUDE -0.1
\RAHMS 13 42 30 \DECDM -62 23
\SIZE 30\X/20? \TYPE C?
\FLUX1GHZ 15? \ALPHA 0.4?
\NAMES

\DATE 14 Dec 2022

\LONGITUDE 309.2 \LATITUDE -0.6
\RAHMS 13 46 31 \DECDM -62 54
\SIZE 15\X/12 \TYPE S
\FLUX1GHZ 7? \ALPHA 0.4?
\NAMES

\DATE 9 Dec 2022

\LONGITUDE 309.8 \LATITUDE +0.0
\RAHMS 13 50 30 \DECDM -62 05
\SIZE 25\X/19 \TYPE S
\FLUX1GHZ 17 \ALPHA 0.5
\NAMES

\DATE 14 Dec 2022

\LONGITUDE 310.6 \LATITUDE -1.6
\RAHMS 14 00 45 \DECDM -63 26
\SIZE 2.5 \TYPE C?
\FLUX1GHZ ? \ALPHA ?
\NAMES

\DATE 9 Dec 2022

\LONGITUDE 310.6 \LATITUDE -0.3
\RAHMS 13 58 00 \DECDM -62 09
\SIZE 8 \TYPE S
\FLUX1GHZ 5? \ALPHA ?
\NAMES Kes 20B

\DATE 9 Apr 2019

\LONGITUDE 310.8 \LATITUDE -0.4
\RAHMS 14 00 00 \DECDM -62 17
\SIZE 12 \TYPE S
\FLUX1GHZ 6? \ALPHA ?
\NAMES Kes 20A

\DATE 13 Dec 2022

\LONGITUDE 311.5 \LATITUDE -0.3
\RAHMS 14 05 38 \DECDM -61 58
\SIZE 5 \TYPE S
\FLUX1GHZ 3? \ALPHA 0.5
\NAMES

\DATE 13 Dec 2022

\LONGITUDE 312.4 \LATITUDE -0.4
\RAHMS 14 13 00 \DECDM -61 44
\SIZE 38 \TYPE S
\FLUX1GHZ 45 \ALPHA 0.36
\NAMES

\DATE 12 Mar 2024

\LONGITUDE 312.5 \LATITUDE -3.0
\RAHMS 14 21 00 \DECDM -64 12
\SIZE 20\X/18 \TYPE S
\FLUX1GHZ 3.5? \ALPHA ?
\NAMES

\DATE 20 Dec 2023

\LONGITUDE 315.1 \LATITUDE +2.7
\RAHMS 14 24 30 \DECDM -57 50
\SIZE 190\X/150 \TYPE S
\FLUX1GHZ ? \ALPHA ?
\NAMES

\DATE 20 May 2014

\LONGITUDE 315.4 \LATITUDE -2.3
\RAHMS 14 43 00 \DECDM -62 30
\SIZE 42 \TYPE S
\FLUX1GHZ 49 \ALPHA 0.6
\NAMES RCW 86, MSH 14$-$6{\sl 3}

\DATE 12 Mar 2024

\LONGITUDE 315.4 \LATITUDE -0.3
\RAHMS 14 35 55 \DECDM -60 36
\SIZE 24\X/13 \TYPE ?
\FLUX1GHZ 8 \ALPHA 0.4
\NAMES

\DATE 14 Dec 2022

\LONGITUDE 315.9 \LATITUDE -0.0
\RAHMS 14 38 25 \DECDM -60 11
\SIZE 25\X/14 \TYPE S
\FLUX1GHZ 0.8? \ALPHA ?
\NAMES

\DATE 14 Dec 2022

\LONGITUDE 316.3 \LATITUDE -0.0
\RAHMS 14 41 30 \DECDM -60 00
\SIZE 29\X/14 \TYPE S
\FLUX1GHZ 20? \ALPHA 0.4
\NAMES (MSH 14$-$5{\sl 7})

\DATE 14 Dec 2022

\LONGITUDE 317.3 \LATITUDE -0.2
\RAHMS 14 49 40 \DECDM -59 46
\SIZE 11 \TYPE S
\FLUX1GHZ 4.7? \ALPHA ?
\NAMES

\DATE 16 Dec 2022

\LONGITUDE 318.2 \LATITUDE +0.1
\RAHMS 14 54 50 \DECDM -59 04
\SIZE 40\X/35 \TYPE S
\FLUX1GHZ $>$3.9? \ALPHA ?
\NAMES

\DATE 14 Dec 2022

\LONGITUDE 318.9 \LATITUDE +0.4
\RAHMS 14 58 30 \DECDM -58 29
\SIZE 30\X/14 \TYPE C
\FLUX1GHZ 4? \ALPHA 0.2?
\NAMES

\DATE 14 Dec 2022

\LONGITUDE 320.4 \LATITUDE -1.2
\RAHMS 15 14 30 \DECDM -59 08
\SIZE 35 \TYPE C
\FLUX1GHZ 60? \ALPHA 0.4
\NAMES MSH 15$-$5{\sl 2}, RCW 89

\DATE 15 Dec 2022

\LONGITUDE 320.6 \LATITUDE -1.6
\RAHMS 15 17 50 \DECDM -59 16
\SIZE 60\X/30 \TYPE S
\FLUX1GHZ ? \ALPHA ?
\NAMES

\DATE 14 Dec 2022

\LONGITUDE 321.9 \LATITUDE -1.1
\RAHMS 15 23 45 \DECDM -58 13
\SIZE 28 \TYPE S
\FLUX1GHZ $>$3.4? \ALPHA ?
\NAMES

\DATE 14 Dec 2022

\LONGITUDE 321.9 \LATITUDE -0.3
\RAHMS 15 20 40 \DECDM -57 34
\SIZE 31\X/23 \TYPE S
\FLUX1GHZ 13 \ALPHA 0.3
\NAMES

\DATE 14 Dec 2022

\LONGITUDE 322.1 \LATITUDE +0.0
\RAHMS 15 20 49 \DECDM -57 10
\SIZE 8\X/4.5? \TYPE S?
\FLUX1GHZ ? \ALPHA ?
\NAMES

\DATE 9 Dec 2022

\LONGITUDE 322.5 \LATITUDE -0.1
\RAHMS 15 23 23 \DECDM -57 06
\SIZE 15 \TYPE C
\FLUX1GHZ 1.5 \ALPHA 0.4
\NAMES

\DATE 13 Aug 1998

\LONGITUDE 323.5 \LATITUDE +0.1
\RAHMS 15 28 42 \DECDM -56 21
\SIZE 13 \TYPE S
\FLUX1GHZ 3? \ALPHA 0.4?
\NAMES

\DATE 16 Mar 2009

\LONGITUDE 323.7 \LATITUDE -1.0
\RAHMS 15 34 30 \DECDM -57 12
\SIZE 51\X/38 \TYPE S
\FLUX1GHZ ? \ALPHA ?
\NAMES

\DATE 12 Mar 2024

\LONGITUDE 326.3 \LATITUDE -1.8
\RAHMS 15 53 00 \DECDM -56 10
\SIZE 38 \TYPE C
\FLUX1GHZ 145 \ALPHA varies
\NAMES MSH 15$-$5{\sl 6}

\DATE 14 Mar 2024

\LONGITUDE 327.1 \LATITUDE -1.1
\RAHMS 15 54 25 \DECDM -55 09
\SIZE 18 \TYPE C
\FLUX1GHZ 7 \ALPHA ?
\NAMES

\DATE 14 Mar 2024

\LONGITUDE 327.2 \LATITUDE -0.1
\RAHMS 15 50 55 \DECDM -54 18
\SIZE 5 \TYPE S
\FLUX1GHZ 0.5 \ALPHA ?
\NAMES

\DATE 14 Mar 2024

\LONGITUDE 327.4 \LATITUDE +0.4
\RAHMS 15 48 20 \DECDM -53 49
\SIZE 21 \TYPE S
\FLUX1GHZ 26 \ALPHA 0.6
\NAMES Kes 27

\DATE 14 Mar 2024

\LONGITUDE 327.4 \LATITUDE +1.0
\RAHMS 15 46 48 \DECDM -53 20
\SIZE 14 \TYPE S
\FLUX1GHZ 1.9 \ALPHA ?
\NAMES

\DATE 14 Mar 2024

\LONGITUDE 327.6 \LATITUDE +14.6
\RAHMS 15 02 50 \DECDM -41 56
\SIZE 30 \TYPE S
\FLUX1GHZ 19 \ALPHA 0.6
\NAMES SN1006, PKS 1459$-$41

\DATE 12 Mar 2024

\LONGITUDE 328.4 \LATITUDE +0.2
\RAHMS 15 55 30 \DECDM -53 17
\SIZE 5 \TYPE F
\FLUX1GHZ 15 \ALPHA 0.0
\NAMES (MSH 15$-$5{\sl 7})

\DATE 14 Mar 2024

\LONGITUDE 329.7 \LATITUDE +0.4
\RAHMS 16 01 20 \DECDM -52 18
\SIZE 40\X/33 \TYPE S
\FLUX1GHZ $>$34? \ALPHA ?
\NAMES

\DATE 14 Dec 2022

\LONGITUDE 330.0 \LATITUDE +15.0
\RAHMS 15 10 00 \DECDM -40 00
\SIZE 180? \TYPE S
\FLUX1GHZ 350? \ALPHA 0.5?
\NAMES Lupus Loop

\DATE 30 May 2014

\LONGITUDE 330.2 \LATITUDE +1.0
\RAHMS 16 01 06 \DECDM -51 34
\SIZE 11 \TYPE S?
\FLUX1GHZ 5? \ALPHA 0.3
\NAMES

\DATE 15 Dec 2022

\LONGITUDE 332.0 \LATITUDE +0.2
\RAHMS 16 13 17 \DECDM -50 53
\SIZE 12 \TYPE S
\FLUX1GHZ 8? \ALPHA 0.5
\NAMES

\DATE 13 May 2017

\LONGITUDE 332.4 \LATITUDE -0.4
\RAHMS 16 17 33 \DECDM -51 02
\SIZE 10 \TYPE S
\FLUX1GHZ 28 \ALPHA 0.5
\NAMES RCW 103

\DATE 12 Mar 2024

\LONGITUDE 332.4 \LATITUDE +0.1
\RAHMS 16 15 20 \DECDM -50 42
\SIZE 15 \TYPE S
\FLUX1GHZ 26 \ALPHA 0.5
\NAMES MSH 16$-$5{\sl 1}, Kes 32

\DATE 30 May 2014

\LONGITUDE 332.5 \LATITUDE -5.6
\RAHMS 16 43 20 \DECDM -54 30
\SIZE 35 \TYPE S
\FLUX1GHZ 2? \ALPHA 0.7?
\NAMES

\DATE 20 Dec 2023

\LONGITUDE 335.2 \LATITUDE +0.1
\RAHMS 16 27 45 \DECDM -48 47
\SIZE 21 \TYPE S
\FLUX1GHZ 16 \ALPHA 0.5
\NAMES

\DATE 14 Dec 2022

\LONGITUDE 336.7 \LATITUDE +0.5
\RAHMS 16 32 11 \DECDM -47 19
\SIZE 14\X/10 \TYPE S
\FLUX1GHZ 6 \ALPHA 0.5
\NAMES

\DATE 25 Apr 2014

\LONGITUDE 337.0 \LATITUDE -0.1
\RAHMS 16 35 57 \DECDM -47 36
\SIZE 1.5 \TYPE S
\FLUX1GHZ 1.5 \ALPHA 0.6?
\NAMES (CTB 33)

\DATE 23 May 2014

\LONGITUDE 337.2 \LATITUDE -0.7
\RAHMS 16 39 28 \DECDM -47 51
\SIZE 6 \TYPE S
\FLUX1GHZ 1.5 \ALPHA 0.4
\NAMES

\DATE 13 May 2017

\LONGITUDE 337.2 \LATITUDE +0.1
\RAHMS 16 35 55 \DECDM -47 20
\SIZE 3\X/2 \TYPE ?
\FLUX1GHZ 1.5? \ALPHA ?
\NAMES

\DATE 19 Feb 2009

\LONGITUDE 337.3 \LATITUDE +1.0
\RAHMS 16 32 39 \DECDM -46 36
\SIZE 15\X/12 \TYPE S
\FLUX1GHZ 16 \ALPHA 0.55
\NAMES Kes 40

\DATE 13 Aug 1998

\LONGITUDE 337.8 \LATITUDE -0.1
\RAHMS 16 39 01 \DECDM -46 59
\SIZE 9\X/6 \TYPE S
\FLUX1GHZ 15 \ALPHA 0.5
\NAMES Kes 41

\DATE 13 Dec 2022

\LONGITUDE 338.1 \LATITUDE +0.4
\RAHMS 16 37 59 \DECDM -46 24
\SIZE 15? \TYPE S
\FLUX1GHZ 4? \ALPHA 0.4
\NAMES

\DATE 28 Apr 2017

\LONGITUDE 338.3 \LATITUDE -0.0
\RAHMS 16 41 00 \DECDM -46 34
\SIZE 8 \TYPE C?
\FLUX1GHZ 7? \ALPHA ?
\NAMES

\DATE 12 Mar 2024

\LONGITUDE 338.5 \LATITUDE +0.1
\RAHMS 16 41 09 \DECDM -46 19
\SIZE 9 \TYPE ?
\FLUX1GHZ 12? \ALPHA ?
\NAMES

\DATE 16 Dec 2022

\LONGITUDE 340.4 \LATITUDE +0.4
\RAHMS 16 46 31 \DECDM -44 39
\SIZE 10\X/7 \TYPE S
\FLUX1GHZ 5 \ALPHA 0.4
\NAMES

\DATE 25 Apr 2014

\LONGITUDE 340.6 \LATITUDE +0.3
\RAHMS 16 47 41 \DECDM -44 34
\SIZE 6 \TYPE S
\FLUX1GHZ 5? \ALPHA 0.4?
\NAMES

\DATE 13 Dec 2022

\LONGITUDE 341.2 \LATITUDE +0.9
\RAHMS 16 47 35 \DECDM -43 47
\SIZE 22\X/16 \TYPE C
\FLUX1GHZ 1.5? \ALPHA 0.6?
\NAMES

\DATE 14 Dec 2022

\LONGITUDE 341.9 \LATITUDE -0.3
\RAHMS 16 55 01 \DECDM -44 01
\SIZE 7 \TYPE S
\FLUX1GHZ 2.5 \ALPHA 0.5
\NAMES

\DATE 27 Jan 2004

\LONGITUDE 342.0 \LATITUDE -0.2
\RAHMS 16 54 50 \DECDM -43 53
\SIZE 12\X/9 \TYPE S
\FLUX1GHZ 3.5? \ALPHA 0.4?
\NAMES

\DATE 1 Aug 2000

\LONGITUDE 342.1 \LATITUDE +0.9
\RAHMS 16 50 43 \DECDM -43 04
\SIZE 10\X/9 \TYPE S
\FLUX1GHZ 0.5? \ALPHA ?
\NAMES

\DATE 13 Aug 1998

\LONGITUDE 343.0 \LATITUDE -6.0
\RAHMS 17 25 00 \DECDM -46 30
\SIZE 250 \TYPE S
\FLUX1GHZ ? \ALPHA ?
\NAMES RCW 114

\DATE 20 Dec 2023

\LONGITUDE 343.1 \LATITUDE -2.3
\RAHMS 17 08 00 \DECDM -44 16
\SIZE 32? \TYPE C?
\FLUX1GHZ 8? \ALPHA 0.5?
\NAMES

\DATE 12 Mar 2024

\LONGITUDE 343.1 \LATITUDE -0.7
\RAHMS 17 00 25 \DECDM -43 14
\SIZE 27\X/21 \TYPE S
\FLUX1GHZ 7.8 \ALPHA 0.55
\NAMES

\DATE 1 Aug 2000

\LONGITUDE 344.7 \LATITUDE -0.1
\RAHMS 17 03 51 \DECDM -41 42
\SIZE 8 \TYPE C?
\FLUX1GHZ 2.5? \ALPHA 0.3?
\NAMES

\DATE 11 Jan 2024

\LONGITUDE 345.1 \LATITUDE -0.2
\RAHMS 17 05 21 \DECDM -41 26
\SIZE 6 \TYPE S
\FLUX1GHZ 1.4? \ALPHA 0.7?
\NAMES

\DATE 4 Jan 2024

\LONGITUDE 345.1 \LATITUDE +0.2
\RAHMS 17 03 40 \DECDM -41 05
\SIZE 10 \TYPE S
\FLUX1GHZ 0.6? \ALPHA 0.6?
\NAMES

\DATE 4 Jan 2024

\LONGITUDE 345.7 \LATITUDE -0.2
\RAHMS 17 07 20 \DECDM -40 53
\SIZE 6 \TYPE S
\FLUX1GHZ 0.6? \ALPHA ?
\NAMES

\DATE 13 Aug 1998

\LONGITUDE 346.6 \LATITUDE -0.2
\RAHMS 17 10 19 \DECDM -40 11
\SIZE 8 \TYPE S
\FLUX1GHZ 8? \ALPHA 0.5?
\NAMES

\DATE 16 Dec 2022

\LONGITUDE 347.3 \LATITUDE -0.5
\RAHMS 17 13 50 \DECDM -39 45
\SIZE 65\X/55 \TYPE S?
\FLUX1GHZ 30? \ALPHA ?
\NAMES RX J1713.7$-$3946

\DATE 16 Dec 2022

\LONGITUDE 348.5 \LATITUDE -0.0
\RAHMS 17 15 26 \DECDM -38 28
\SIZE 10? \TYPE S?
\FLUX1GHZ 10? \ALPHA 0.4?
\NAMES

\DATE 13 Dec 2022

\LONGITUDE 348.5 \LATITUDE +0.1
\RAHMS 17 14 06 \DECDM -38 32
\SIZE 15 \TYPE S
\FLUX1GHZ 72 \ALPHA 0.3
\NAMES CTB 37A

\DATE 13 Dec 2022

\LONGITUDE 348.7 \LATITUDE +0.3
\RAHMS 17 13 55 \DECDM -38 11
\SIZE 17? \TYPE S
\FLUX1GHZ 26 \ALPHA 0.3
\NAMES CTB 37B

\DATE 9 Dec 2022

\LONGITUDE 348.8 \LATITUDE +1.1
\RAHMS 17 11 29 \DECDM -37 36
\SIZE 10 \TYPE S
\FLUX1GHZ 0.6? \ALPHA 0.7?
\NAMES

\DATE 20 Dec 2022

\LONGITUDE 349.2 \LATITUDE -0.1
\RAHMS 17 17 15 \DECDM -38 04
\SIZE 9\X/6 \TYPE S
\FLUX1GHZ 1.4? \ALPHA ?
\NAMES

\DATE 27 Aug 1996

\LONGITUDE 349.7 \LATITUDE +0.2
\RAHMS 17 17 59 \DECDM -37 26
\SIZE 2.5\X/2 \TYPE S
\FLUX1GHZ 20 \ALPHA 0.5
\NAMES

\DATE 13 Dec 2022

\LONGITUDE 350.0 \LATITUDE -2.0
\RAHMS 17 27 50 \DECDM -38 32
\SIZE 45 \TYPE S
\FLUX1GHZ 26 \ALPHA 0.4
\NAMES

\DATE 13 May 2017

\LONGITUDE 350.1 \LATITUDE -0.3
\RAHMS 17 21 05 \DECDM -37 27
\SIZE 4? \TYPE ?
\FLUX1GHZ 6? \ALPHA 0.8?
\NAMES

\DATE 16 Dec 2022

\LONGITUDE 351.0 \LATITUDE -5.4
\RAHMS 17 46 00 \DECDM -39 25
\SIZE 30 \TYPE S
\FLUX1GHZ ? \ALPHA ?
\NAMES

\DATE 12 Jun 2017

\LONGITUDE 351.2 \LATITUDE +0.1
\RAHMS 17 22 27 \DECDM -36 11
\SIZE 7 \TYPE C?
\FLUX1GHZ 5? \ALPHA 0.4
\NAMES

\DATE 13 Dec 2022

\LONGITUDE 351.7 \LATITUDE +0.8
\RAHMS 17 21 00 \DECDM -35 27
\SIZE 18\X/14 \TYPE S
\FLUX1GHZ 10 \ALPHA 0.5?
\NAMES

\DATE 14 Dec 2022

\LONGITUDE 351.9 \LATITUDE -0.9
\RAHMS 17 28 52 \DECDM -36 16
\SIZE 12\X/9 \TYPE S
\FLUX1GHZ 1.8? \ALPHA ?
\NAMES

\DATE 12 Mar 2024

\LONGITUDE 352.7 \LATITUDE -0.1
\RAHMS 17 27 40 \DECDM -35 07
\SIZE 8\X/6 \TYPE S
\FLUX1GHZ 4 \ALPHA 0.6
\NAMES

\DATE 12 Mar 2024

\LONGITUDE 353.3 \LATITUDE -1.1
\RAHMS 17 33 10 \DECDM -35 12
\SIZE 60 \TYPE S
\FLUX1GHZ 24? \ALPHA 0.85?
\NAMES

\DATE 20 Dec 2022

\LONGITUDE 353.6 \LATITUDE -0.7
\RAHMS 17 32 00 \DECDM -34 44
\SIZE 30 \TYPE S
\FLUX1GHZ 2.5? \ALPHA ?
\NAMES

\DATE 12 Mar 2024

\LONGITUDE 353.9 \LATITUDE -2.0
\RAHMS 17 38 55 \DECDM -35 11
\SIZE 13 \TYPE S
\FLUX1GHZ 1? \ALPHA 0.5?
\NAMES

\DATE 8 Nov 2001

\LONGITUDE 354.1 \LATITUDE +0.1
\RAHMS 17 30 28 \DECDM -33 46
\SIZE 15\X/3? \TYPE C?
\FLUX1GHZ ? \ALPHA varies
\NAMES

\DATE 4 Jun 2017

\LONGITUDE 354.8 \LATITUDE -0.8
\RAHMS 17 36 00 \DECDM -33 42
\SIZE 19 \TYPE S
\FLUX1GHZ 2.8? \ALPHA ?
\NAMES

\DATE 1 Aug 2000

\LONGITUDE 355.4 \LATITUDE +0.7
\RAHMS 17 31 20 \DECDM -32 26
\SIZE 25 \TYPE S
\FLUX1GHZ 5? \ALPHA ?
\NAMES

\DATE 14 Dec 2022

\LONGITUDE 355.6 \LATITUDE -0.0
\RAHMS 17 35 16 \DECDM -32 38
\SIZE 8\X/6 \TYPE S
\FLUX1GHZ 3? \ALPHA ?
\NAMES

\DATE 25 Apr 2014

\LONGITUDE 355.9 \LATITUDE -2.5
\RAHMS 17 45 53 \DECDM -33 43
\SIZE 13 \TYPE S
\FLUX1GHZ 8 \ALPHA 0.5
\NAMES

\DATE 25 Apr 2014

\LONGITUDE 356.2 \LATITUDE +4.5
\RAHMS 17 19 00 \DECDM -29 40
\SIZE 25 \TYPE S
\FLUX1GHZ 4 \ALPHA 0.7
\NAMES

\DATE 20 Dec 2023

\LONGITUDE 356.3 \LATITUDE -1.5
\RAHMS 17 42 35 \DECDM -32 52
\SIZE 20\X/15 \TYPE S
\FLUX1GHZ 3? \ALPHA ?
\NAMES

\DATE 20 May 2014

\LONGITUDE 356.3 \LATITUDE -0.3
\RAHMS 17 37 56 \DECDM -32 16
\SIZE 11\X/7 \TYPE S
\FLUX1GHZ 3? \ALPHA ?
\NAMES

\DATE 13 May 2017

\LONGITUDE 357.7 \LATITUDE -0.1
\RAHMS 17 40 29 \DECDM -30 58
\SIZE 8\X/3? \TYPE ?
\FLUX1GHZ 37 \ALPHA 0.4
\NAMES MSH 17$-$3{\sl 9}

\DATE 16 Dec 2022

\LONGITUDE 357.7 \LATITUDE +0.3
\RAHMS 17 38 35 \DECDM -30 44
\SIZE 24 \TYPE S
\FLUX1GHZ 10 \ALPHA 0.4?
\NAMES

\DATE 14 Dec 2022

\LONGITUDE 358.0 \LATITUDE +3.8
\RAHMS 17 26 00 \DECDM -28 36
\SIZE 38 \TYPE S
\FLUX1GHZ 1.5? \ALPHA ?
\NAMES

\DATE 20 Dec 2023

\LONGITUDE 358.1 \LATITUDE +1.0
\RAHMS 17 37 00 \DECDM -29 59
\SIZE 20 \TYPE S
\FLUX1GHZ 2? \ALPHA ?
\NAMES

\DATE 20 Dec 2022

\LONGITUDE 358.5 \LATITUDE -0.9
\RAHMS 17 46 10 \DECDM -30 40
\SIZE 17 \TYPE S
\FLUX1GHZ 4? \ALPHA ?
\NAMES

\DATE 20 Dec 2022

\LONGITUDE 359.0 \LATITUDE -0.9
\RAHMS 17 46 50 \DECDM -30 16
\SIZE 23 \TYPE S
\FLUX1GHZ 23 \ALPHA 0.5
\NAMES

\DATE 20 Dec 2022

\LONGITUDE 359.1 \LATITUDE -0.5
\RAHMS 17 45 30 \DECDM -29 57
\SIZE 24 \TYPE S
\FLUX1GHZ 14 \ALPHA 0.4?
\NAMES

\DATE 4 Jan 2024

\LONGITUDE 359.1 \LATITUDE +0.9
\RAHMS 17 39 36 \DECDM -29 11
\SIZE 12\X/11 \TYPE S
\FLUX1GHZ 2? \ALPHA ?
\NAMES

\DATE 20 Dec 2022

\LONGITUDE 359.2 \LATITUDE -1.1
\RAHMS 17 48 14 \DECDM -30 12
\SIZE 5\X/4 \TYPE S?
\FLUX1GHZ 0.4? \ALPHA 1.1?
\NAMES

\DATE 20 Dec 2022

\endsnrcat

\end{document}